# Viscosity, breakdown of Stokes-Einstein relation and dynamical heterogeneity in supercooled liquid $Ge_2Sb_2Te_5$ from simulations with a neural network potential

Simone Marcorini,[1] Rocco Pomodoro,[1] Omar Abou El Kheir,[1] and Marco Bernasconi[1]
*Department of Materials Science, University of Milano-Bicocca, Via R. Cozzi 55, I-20125 Milano, Italy*

(*Electronic mail: marco-bernasconi@unimib.it)

Phase change materials are exploited in non-volatile electronic memories and photonic devices that rely on a fast and reversible transformation between the amorphous and crystalline phase upon heating. The recrystallization of the amorphous phase at the operation conditions of the memories occurs in the supercooled liquid phase above the glass transition temperature $T_g$. The dynamics of the supercooled liquid is thus of great relevance for the operation of the devices and, close to $T_g$, also for the structural relaxations of the glass that affect the performances of the memories. Information on the atomic dynamics is provided by the diffusion coefficient ($D$) and by the viscosity ($\eta$) which are, however, both difficult to be measured experimentally at the operation conditions of the devices due to the fast crystallization. In this work, we leverage a machine learning interatomic potential for the flagship phase change compound $Ge_2Sb_2Te_5$ to compute $\eta$, $D$ and the $\alpha$-relaxation time in a wide temperature range from 1200 K to about 100 K above $T_g$. Large scale molecular dynamics simulations allowed quantifying the fragility of the liquid and the occurrence of a breakdown of the Stokes-Einstein relation between $\eta$ and $D$ in the supercooled phase. Isoconfigurational analysis provided a visualization of the emergence of dynamical heterogeneities responsible for the breakdown of the Stokes-Einstein relation. The analysis revealed that the regions of most mobile atoms are related to the presence of Ge atoms with particular local environments.

## I. INTRODUCTION

Phase change materials such as the flagship compound $Ge_2Sb_2Te_5$ (GST) are exploited in nonvolatile data storage thanks to their ability to undergo a rapid and reversible transformation between the crystalline and amorphous phases upon heating.[1–5] The large difference in electrical properties between the two phases allows encoding a binary information that can be read by the measurement either of the optical reflectivity in optical media (DVDs and Blu-rays disks) or of the electrical resistivity in the electronic phase change memories (PCMs). The phase transformation is induced by laser radiation in the optical disks and by Joule heating via current pulses in PCMs. Partial recrystallization in the set operation or modulation of the size of the amorphous region in the reset operation can also be exploited to implement analogical resistance states for applications in in-memory and neuromorphic computing.[6]

A key functional property for these applications is the ability of GST to undergo a rapid crystallization (10-100 ns) in the supercooled liquid phase around 550-700 K and at the same time to be highly stable in the amorphous phase just below the glass transition temperature $T_g$ (473 K from the most recent work[7]) to guarantee data retention. This feature is made possible by the fragility of the supercooled liquid which leads to a huge change in the atomic mobility (self-diffusion coefficient $D$) in a relatively narrow temperature range above $T_g$. According to the taxonomy introduced by Angell,[8] supercooled liquids are classified as strong or fragile on the basis of the temperature dependence of their viscosity $\eta$. By definition, a strong liquid displays an Arrhenius behavior of $\eta$ from the melting temperature $T_m$ down to $T_g$. In contrast, a fragile liquid features a super-Arrhenius behavior of $\eta$ which keeps very low from $T_m$ down to low temperatures where a steep rise is then observed to reach the high value of $10^{12}$ Pa s expected at $T_g$.[8] The degree of fragility is quantified by the fragility index $m$ defined by $m = d(\log_{10}\eta)/d(T_g/T)\mid_{T=T_g}$, which measures the slope of $\eta$ as a function of $T_g/T$ evaluated at the glass transition temperature. Due to the Stokes-Einstein relation (SER) between $\eta$ and $D$ ($\eta = k_B T/(6\pi R D)$, where $R$ is the average van der Waals radius), in the fragile liquid the super-Arrhenius behavior is observed as well in the diffusion coefficient $D$. In turns, $D$ enters as a kinetic prefactor both in the crystal growth velocity $v_g \propto D(1-\exp(-\Delta\mu/(k_B T))$, where $\Delta\mu$ is the thermodynamical driving force for the crystallization corresponding to the difference in the free energy between the liquid and the crystal, and in the crystal nucleation rate $I_{ss} \propto D\exp(-G_c/(k_B T))$, where $G_c$ is the free energy of formation of the overcritical crystal nucleus which also depends on $\Delta\mu$ as $G_c = 16\pi\sigma^3/(3\rho^2\Delta\mu^2)$ within classical nucleation theory,[9] where $\sigma$ is the crystal-liquid interface energy and $\rho$ is the crystal density. $\Delta\mu$ is zero at $T_m$ and increases by decreasing temperature favoring crystallization at high supercooling. In a fragile liquid, $D$ keeps high down to temperature close to $T_g$ which allows having both a high kinetic prefactor and a high thermodynamical driving force for nucleation and growth. On the other hand, a high fragility is also responsible for larger structural relaxations below $T_g$[10] which leads to a drift in electrical resistance with time.[11] This drift is detrimental for the operation of the devices, especially for neuromorphic applications for which several different resistance levels must be resolved over time. This problem can be mitigated by a suitable design of the cell architecture.[12] Anyway, it would be desirable for a phase change material to feature a fragile-to-strong crossover (FSC) at temperature $T_{fs}$ below which the system becomes strong. Ideally, $T_{fs}$ should be sufficiently close to $T_g$ to still guarantee a high $D$ down to low temperatures to speed up the crystallization rate, but at the same time also to limit structural relaxations in the amorphous phase.

The fast crystallization enabled by the high fragility brings, however, some hindrances to the experimental measurements of $\eta$ itself, and then to the assessment of the presence of a FSC.[13–15] Experimental information on the viscosity was obtained only indirectly from the crystal growth velocity inferred from ultrafast differential scanning calorimetry (DSC). The first ultrafast DSC measurement on thin film yielded a fragility index $m \simeq 90$ and $T_g$=383 K,[16] while more recent measurements yielded much lower values ranging from 40 for thin films,[17] to 57-62 for nanoparticles incorporating $H_2$ or $CH_4$.[18] A different experimental technique based on mechanical analysis of thin films also yielded $m$=47.[19] The low values of $m$ were ascribed to a FSC at $T_{fs}$=$T_g$/0.85 from the analysis of nanoparticles in Ref.[18], and more recently also in thin films where a $T_{fs}$ as large as 683 K was inferred from DSC.[20] A more recent DSC work has shown,[7] however, that crystallization takes place in the amorphous phase at low heating rates and in the supercooled liquid at high heating rates. Only at the highest heating rates the endothermic peak at $T_g$ was disentangled from the exothermic peak due to crystallization. This analysis yielded a $T_g$ of about 473 K,[7] which is higher than the values previously used for $T_g$ that was usually identified with the crystallization temperature. Therefore, as discussed in Ref.[13], deriving viscosity from crystallization rate is not straightforward which still raises some uncertainties on the evaluation of $m$ and $T_{fs}$ in GST.

On the other hand, molecular dynamics (MD) simulations represent an alternative tool to explore the dynamics of supercooled liquids which can provide an independent evaluation of viscosity, diffusion coefficient and crystal growth velocities to establish a correlation among these quantities.[2,21] Simulations based on density functional theory (DFT) have been performed at high temperatures in the liquid above $T_m$,[22,23] or in the supercooled phase down to 820 K,[24] to compute $\eta$ from $D$ and the application of SER or from the relaxation time $\tau_\alpha$ obtained from the intermediate scattering function (see Sec. II) and the application of the Maxwell relation ($\eta = G_\infty \tau_\alpha$).[24] At low temperatures, however, a breakdown of SER is usually expected in a fragile supercooled liquid which prevents the use of $D$ to estimate $\eta$, and on the other hand limitations in the simulation time prevent to compute directly $\eta$ or $\tau_\alpha$ from DFT-MD. In this respect, we remark that a breakdown of SER was indeed inferred experimentally from DSC below $T_m$,[16] but also above $T_m$ from quasi-elastic neutron scattering.[25]

The limitations of DFT-MD can be overcome by exploiting machine learning techniques that have opened a new avenue in the modeling of materials by providing interatomic potentials for large scale MD simulations from the fitting of a DFT database of energies and forces.[26–28] Concerning phase change materials, a machine-learned interatomic potential (MLIP) based on the neural network (NN) method of Ref.[26] was developed for GeTe in 2012,[29] and then applied to compute the crystal growth velocity,[30,31] the viscosity from the Green-Kubo relation (see Sec. II),[32,33] and to assess the presence of dynamical heterogeneity (DH) responsible for the breakdown of SER.[34] More recently, MLIP for $Ge_2Sb_2Te_5$ have been developed as well.[35–38]

In this work, we exploit the MLIP for $Ge_2Sb_2Te_5$, that we developed in Ref.[36] by using the NN scheme of the DeepMD code,[39–41] to study the viscosity, the breakdown of SER and the emergence of dynamical heterogeneities in the supercooled liquid down to 550 K.

## II. COMPUTATIONAL DETAILS

Molecular dynamics simulations were performed by using the MLIP for GST developed in our previous work[36] by using the DeePMD package.[39–41] The NN was trained on a DFT database of energies and forces of about 180,000 configurations of small supercells (57-108 atoms) computed by employing the Perdew-Burke-Ernzerhof (PBE) exchange and correlation functional[42] and norm conserving pseudopotentials.[43] The potential was validated on the structural and dynamical properties of the liquid, amorphous and crystalline phases and it was exploited to study the crystallization kinetics in the bulk,[36] in heterostructures[44] and in a multimillion-atom model of a PCM cell.[45] MD simulations were performed with the DeePMD code by using the Lammps code as MD driver,[46] a time step of 2 fs, and a Nosé-Hoover thermostat.[47,48] In the present simulations we added van der Waals (vdW) interactions that were shown to bring DFT results on viscosity above $T_m$ (from $D$ and SER) to a closer agreement with experiment in GST[22] and GeTe.[49] Here, we use the Grimme (D2) vdW correction.[50] We also remark that the addition of vdW-D2 corrections brings the theoretical crystal growth velocities of GST very close to the experimental DSC data as shown in our previous work.[44] Simulations were performed with a 3996-atom cubic supercell at different temperatures. Since the experimental density of the amorphous phase at 300 K ($\rho$=0.0309 atom/Å$^3$)[51] is very close to the experimental density of the liquid at $T_m$ (0.03075 atom/Å$^3$ at 873 K),[22] we performed simulations in the supercooled liquid at a fixed density of $\rho$=0.03075 atom/Å$^3$. Above $T_m$ we performed two sets of simulations, one at the same fixed density of $\rho$=0.03075 atom/Å$^3$, and a second by adjusting the density with temperature according to the experimental data in Ref.[22]. We remark that the theoretical melting temperature estimated with the phase coexistence method in Ref.[36] in NN+vdW simulations is 940 K.

We considered six temperatures above $T_m$, namely 950-1000-1050-1100-1150-1200, and 14 temperatures below $T_m$, namely 900-850-800-750-700-675-650-625-600-590-580-570-560-550. The system was initially equilibrated at 900 K for 40 ps and then quenched in 150 ps to the target temperature. Then, NVT simulations lasting from 6.6 to 13.2 ns each have been performed to compute the viscosity. We also performed simulations at constant energy (NVE) to compute the diffusion coefficient. A synoptic table of all simulations performed is shown in Table SI in the supplementary material.

We computed the incoherent (self) intermediate scattering function (ISF) $F_s(q,t)$ and the coherent (distinct) ISF $F_d(q,t)$ defined by[52]

$$F_s(q,t) = \langle \Phi_s(q,t) \rangle \quad (1)$$

$$\Phi_s(q,t) = \frac{1}{N} \sum_{j}^{N} \exp\{i\vec{q}\cdot[\vec{r}_j(t_o) - \vec{r}_j(t+t_o)]\} \quad (2)$$

$$F_d(q,t) = \langle \Phi_d(q,t) \rangle \quad (3)$$

$$\Phi_d(q,t) = \frac{1}{N} \sum_{j\neq k}^{N} \exp\{i\vec{q}\cdot[\vec{r}_k(t_o) - \vec{r}_j(t+t_o)]\} \quad (4)$$

where the sum runs over the $N$ atoms at positions $\vec{r}_j$, $\vec{r}_k$ at different times. The average $\langle \ldots \rangle$ is over the initial times $t_o$. We averaged over an equal number of initial times $t_o$ for all times $t$ to have the same statistical accuracy at all times. Since in a liquid $F_s(q,t)$ depends only on the modulus $q$ of $\vec{q}$, we also averaged over 6 directions of $\vec{q}$. We computed $F_s(q,t)$ and $F_d(q,t)$ at $q = q_o = 2$ Å$^{-1}$ which corresponds to the position of the main peak of the static structure factor measured experimentally in Ref.[22]. $F_s(q_o,t)$ was computed from NVT and also NVE simulations (see Table SI in the supplementary material).

The viscosity $\eta$ has been computed by means of the Green-Kubo formula[53] in NVT simulations

$$\eta = \lim_{t\to\infty} \frac{V}{k_B T} \int_0^t \langle \sigma_{xy}(t'+t_o)\sigma_{xy}(t_o) \rangle dt' = \lim_{t\to\infty} \eta(t) \quad (5)$$

where $V$ is the supercell volume, $T$ is the temperature, and $\sigma_{xy}$ is the off-diagonal component of the stress tensor. The average $\langle \ldots \rangle$ is over the initial times $t_o$. We also took an average over the three off-diagonal components $xy$, $xz$, and $yz$. In practice, $\eta$ is evaluated from a plateau in the value of $\eta(t)$ at intermediate times, as at longer times problems in convergence arise due to accumulation of noise.[54] We choose a maximum $t_{max}$ to evaluate $\eta(t)$ equal to twice the time at which the correlation function $\langle \sigma_{xy}(t+t_o)\sigma_{xy}(t_o) \rangle$ goes to zero. We checked that our estimates of $\eta$ change by at most 5 % by doubling $t_{max}$ at all temperatures. We performed blocks averages by first averaging over $t_o$ in $\langle \sigma_{xy}(t+t_o)\sigma_{xy}(t_o) \rangle$ in blocks with time length of $2t_{max}$. We then obtain the average $\eta(t)$ resulting from Eq. 5 and its mean square errors by averaging over blocks (from 500 at high temperatures to about 20 at the lowest temperatures). The convergence of $\eta(t_{max})$ as a function of the number of blocks is shown in Figure S1 in the supplementary material. At the lowest temperatures when the convergence with the number of blocks is slower, we averaged over blocks from different independent simulations to avoid crystal nucleation that invariably occurs on very long times below $T_m$. The formation of crystalline nuclei was monitored by computing the local order parameter for crystallinity $Q_4^{dot}$,[55,56] that we considered in our previous work on GST.[36] Obviously, only trajectories before crystal nucleation were used for the evaluation of $\eta$.

## III. RESULTS

### A. Diffusion coefficient

We first computed the diffusion coefficient $D$ from the average mean squared displacement $\langle r^2(t) \rangle$ (MSD) and the Einstein relation $\lim_{t\to\infty} \langle r^2(t) \rangle = 6Dt$ in NVE simulations at different temperatures (see Table SI in the supplementary material). $D$ is obtained from the linear slope at long times of the MSD shown in Figure 1a. The resulting diffusion coefficient as a function of temperatures is shown in Figure 1b. A high degree of fragility, to be quantified later, is already apparent from the non-Arrhenius behavior of the diffusion coefficient in Figure 1b. The addition of the vdW interaction to our NN potential leads to a reduction of $D$ up to a factor about five at 550 K with respect to NN (PBE) simulations, as it was discussed in our previous work.[44] The comparison with previous DFT calculations of $D$ in the literature[22,24,57] was discussed in Ref.[36]. Note that below 590 K, the points in Fig. 1b appear to be aligned on a straight line, suggesting the presence of a FSC. We will comment on this issue in sec. C when discussing our results on viscosity. The diffusion coefficient for the different species as a function of temperature is shown in Figure S2 in the supplementary material.

### B. Intermediate scattering function and relaxation times

The incoherent intermediate scattering function $F_s(q_o,t)$ (ISF, see Sec. II) from NVT simulations at different temperatures is shown in Figure 2. At lower temperatures $F_s(q_o,t)$ displays a two-step decay with a plateau at intermediate times that becomes more evident by approaching $T_g$. The relaxation toward the plateau is known as $\beta$-relaxation, while the decay to zero at longer times after the plateau is called $\alpha$-relaxation.[58] The characteristic time of the $\alpha$-relaxation, $\tau_\alpha$, can be defined by fitting the ISF at long times with the Kohlrausch-Williams-Watts (KWW)[59,60] stretched exponential function $F_s(q_o,t) \sim e^{-(\frac{t}{\tau_\alpha})^\beta}$. For temperatures not too low, a similar estimate of $\tau_\alpha$ is obtained by the simple condition $F_s(q_o,t) = 1/e$ as it is shown in Figure S3 in the supplementary material. We note that we chose $q_o$= 2 Å$^{-1}$ that corresponds to the main peak in the experimental static structure factor to measure the relaxation time at the main structural length scale. The relaxation time is in fact dependent on $q$ as shown for three representative temperatures in Figure S4 in the supplementary material. The KWW fit of $F_s(q_o,t)$ is shown in Figure S5 in the supplementary material. The resulting KWW exponent $\beta$ as a function of temperature is shown in Figure 3. The fitting of $F_s(q_o,t)$ resolved for the different species yields a coefficient $\beta$ that are within the error bars of the value obtained from the total ISF. The relaxation time $\tau_\alpha$ obtained from the KWW fit is shown in Figure 4(a)-(b) for temperatures below and above $T_m$ (see Figure S6 in the supplementary material for $\tau_\alpha$ resolved for the different species). Previous results from DFT-MD simulations at high temperatures[24] are also shown in Figure 4(a). Experimentally, the relaxation time was measured above $T_m$ from quasi-elastic neutron scattering at $q_o$ = 2 Å$^{-1}$ which is dominated by coherent scattering.[25] The experimental relaxation times must then be compared with the coherent relaxation time $\tau_\alpha^{cohe}$ obtained from the coherent (distinct) ISF $F_d(q_o,t)$ (see Figure S7 in the supplementary material). The resulting $\tau_\alpha^{cohe}$ obtained from the condition $F_d(q_o,t) = 1/e$ (as usual, at long times $F_d$ is more noisy than $F_s$) for temperatures above T$_m$ are compared with experimental data in Figure 4c.

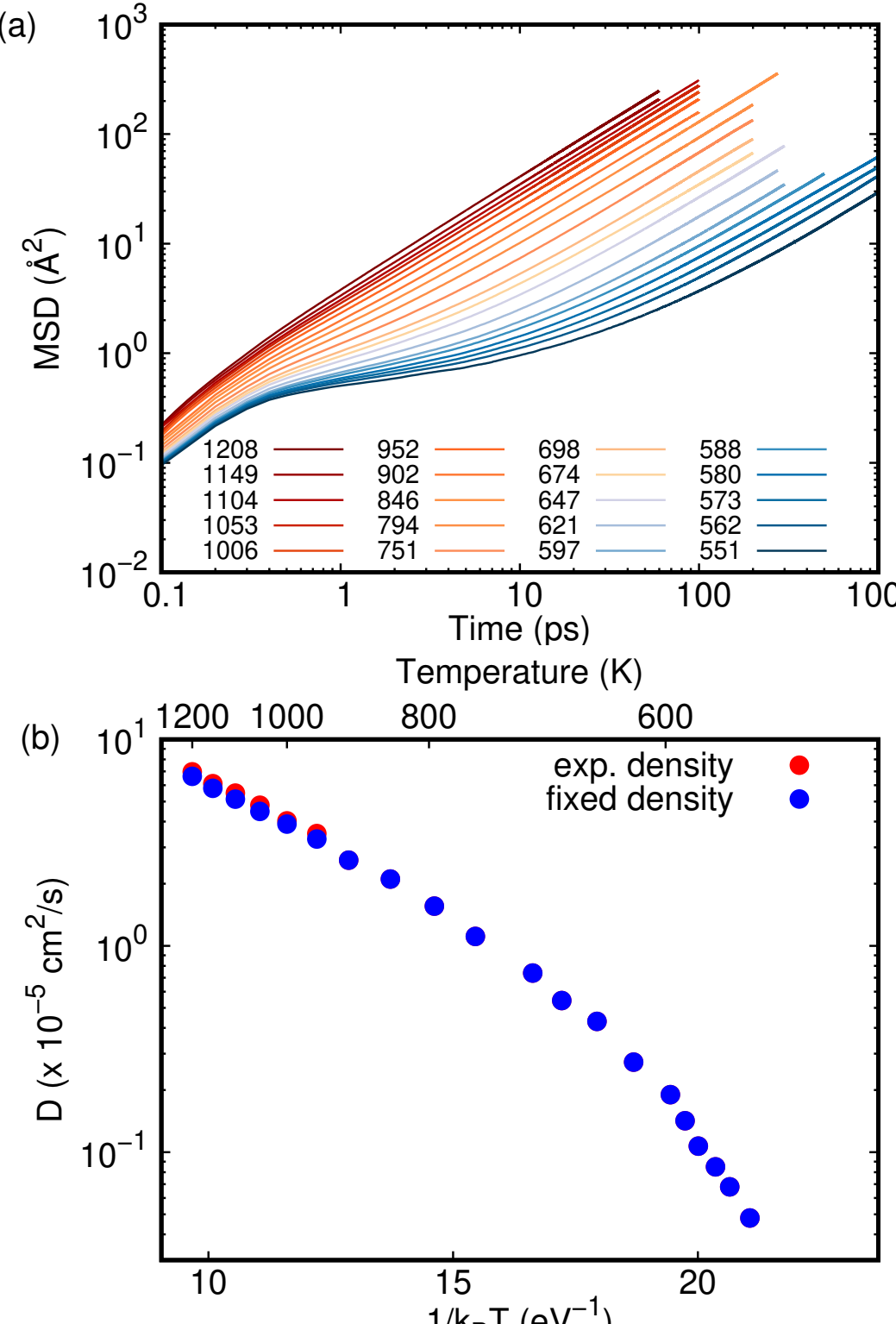

FIG. 1. (a) Mean square displacement (MSD) as a function of time at different temperatures (K) from NVE simulations (b) Diffusion coefficient $D$ as a function of temperature obtained from the MSD and the Einstein relation (see text) from simulations at density fixed to the experimental density at $T_m$ (blue dots) and at the experimental temperature-dependent densities above $T_m$ (red dots).

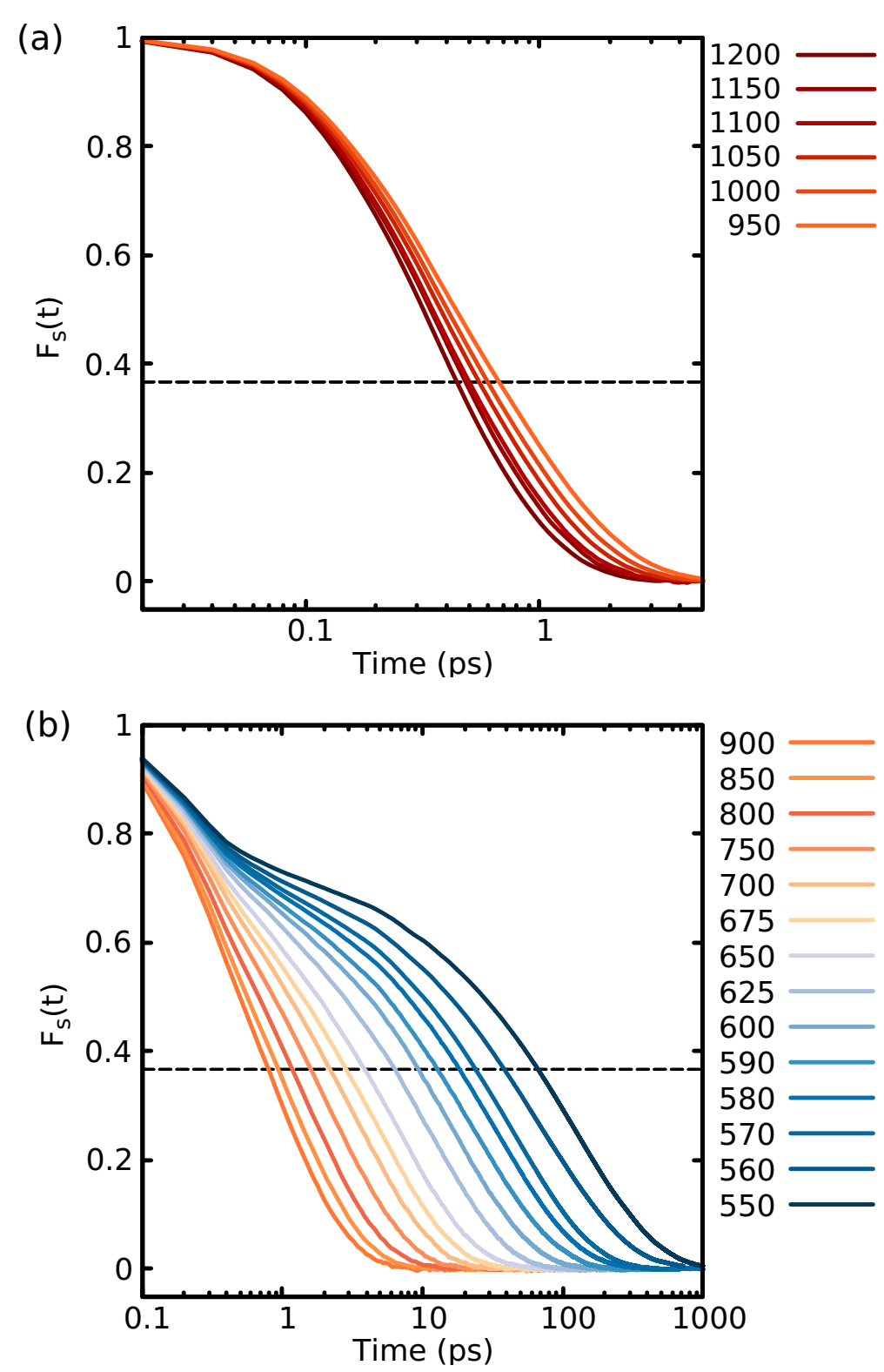

FIG. 2. Incoherent intermediate scattering function $F_s(q_o,t)$ as a function of time at different temperatures (a) above $T_m$ and (b) below $T_m$ from NVT simulations. We chose $q_o$ = 2 Å$^{-1}$ that corresponds to the main peak in the experimental static structure factor (see text). The horizontal line indicates the condition $F_s(q_o,t) = 1/e$.

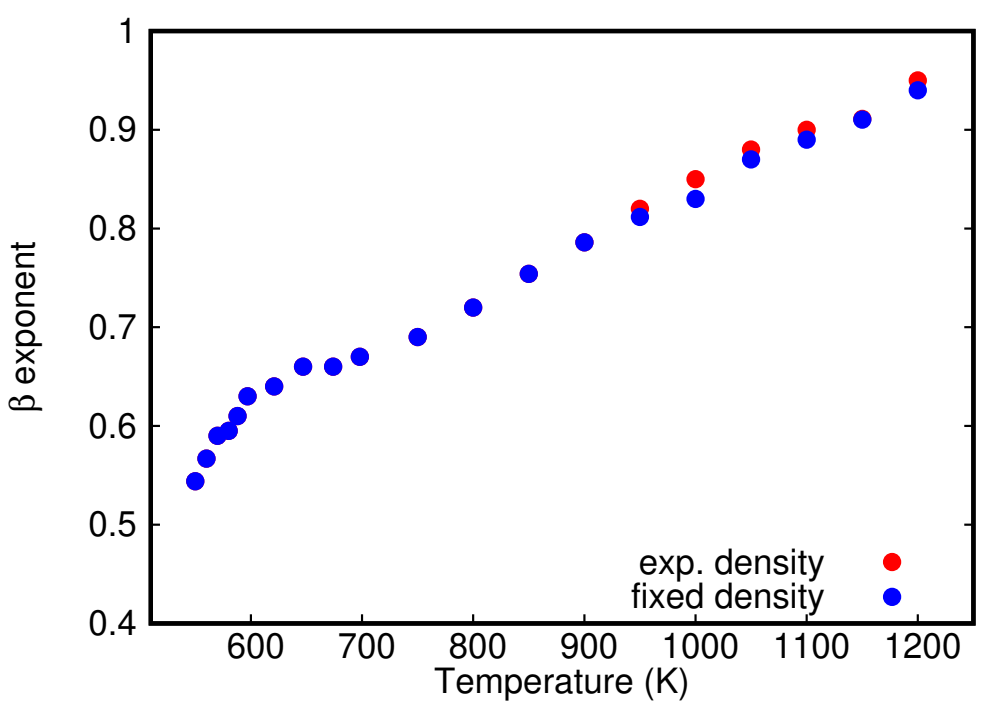

FIG. 3. The KWW exponent $\beta$ (see text) obtained from the fitting of the ISF of Figure 2 as a function of temperature, from simulations at density fixed to the experimental density at $T_m$ (blue dots) and at the experimental temperature-dependent densities above $T_m$ (red dots).

The temperature dependence of $\tau_\alpha$ already shows a high degree of fragility that will be discussed in the next section in connection with the calculation of the viscosity. A nearly linear slope is, however, observed at the lowest temperatures, which mirrors the behavior of $D$ in Figure 1b that will be discussed in the next section. We note that when averaged over simulations of the same length, the $F_s(q_o,t)$ functions obtained from NVT or NVE simulations at the same temperatures are indistinguishable (see Figure S8 in the supplementary material.).

### C. Viscosity and fragility

The viscosity $\eta(t)$ as a function of the integration time $t$ in the GK formula (Eq. 5) is shown in Figure 5 at different temperatures from NVT simulations. The averaging procedure to compute the viscosity is described in Sec. II. The self-correlation function $\langle\sigma_{xy}(t+t_o)\sigma_{xy}(t_o)\rangle$ entering in the GK in-tegral is shown in Figure S9 in the supplementary material for a few representative temperatures. Our estimates for $\eta$ at different temperatures extracted from the plateaus of Figure 5 are reported as an Angell plot in Figure 6 (see also Table SI in the supplementary material). As described in Sec. II, we also repeated the calculation of $\eta$ for $T > T_m$ at the experimental densities. Below $T_m$, the density was always kept fixed because the experimental density of the liquid at $T_m$ and of the amorphous phase at 300 K are very similar (see Sec. II). This new set of data at high temperatures are also collected in Figure 6 (see also Table SI in the supplementary material). Comparison with the experimental data at high temperature from Ref.[22] is given in the inset of Figure 5. The theoretical results are only slightly higher than the experimental values in the whole temperature range above melting. We mention that very recently the viscosity of GST in the liquid and supercooled liquid phase down to about 650 K has been computed by using an ACE machine learning interatomic potential.[38] The resulting viscosity is, however, a factor two lower than in experiments[22] at the experimental melting temperature (1.0 vs 2.01 mPa s).

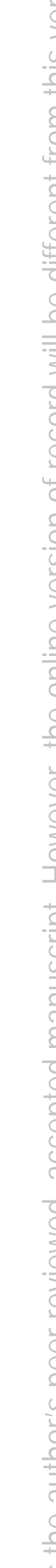

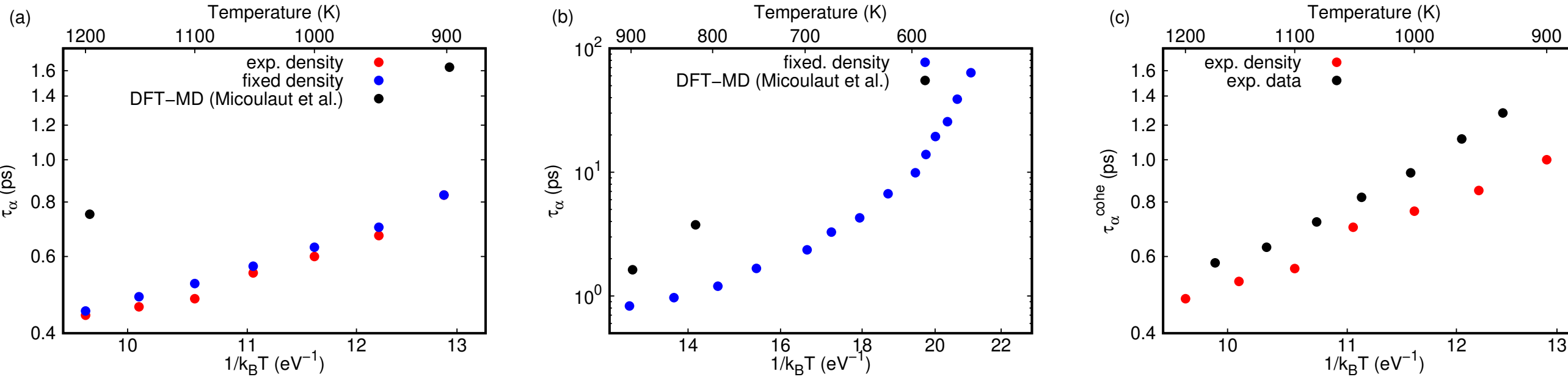

FIG. 4. The relaxation time $\tau_\alpha$ as a function of temperature (a) above $T_m$ and (b) below $T_m$. Previous theoretical results from DFT-MD simulations[24] are shown shown in panel (a)-(b). (c) The coherent relaxation times $\tau_\alpha^{cohe}$ obtained from $F_d(q_o,t)$ at $T > T_m$ (see Figure S7 in the supplementary material) are compared with experimental data (black dots) from quasi-elastic neutron scattering of Ref.[25]. The theoretical data refer to simulations at density fixed to the experimental density at $T_m$ (blue dots) and at the experimental temperature-dependent densities above $T_m$ (red dots).

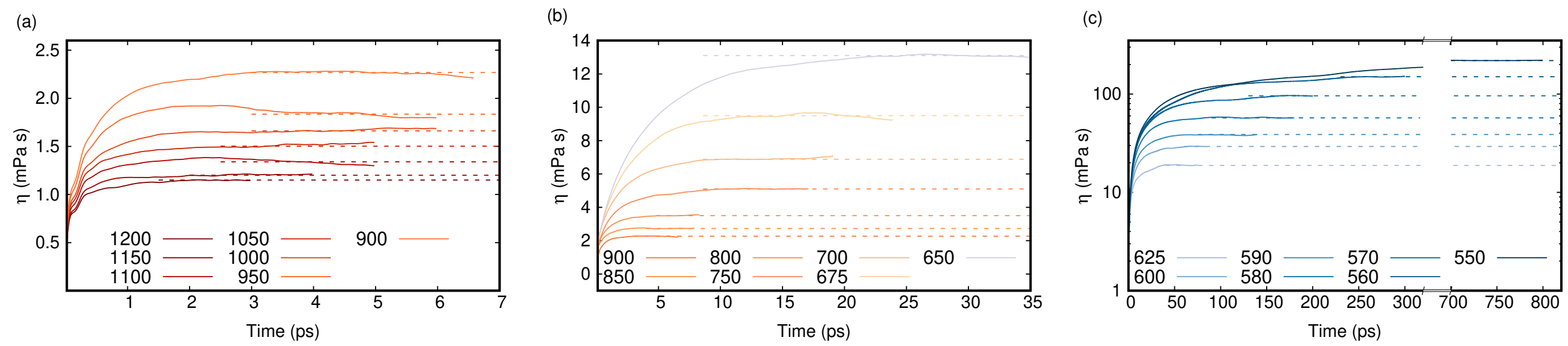

FIG. 5. (a)-(c) Viscosity $\eta(t)$ as a function of the integration time $t$ in the GK formula (Eq. 5) in the three different temperature ranges. The horizontal lines are the averages of $\eta(t)$ over the plateau that gives our estimate of the viscosity $\eta$. The data refer to simulations at density fixed to the experimental density at $T_m$.

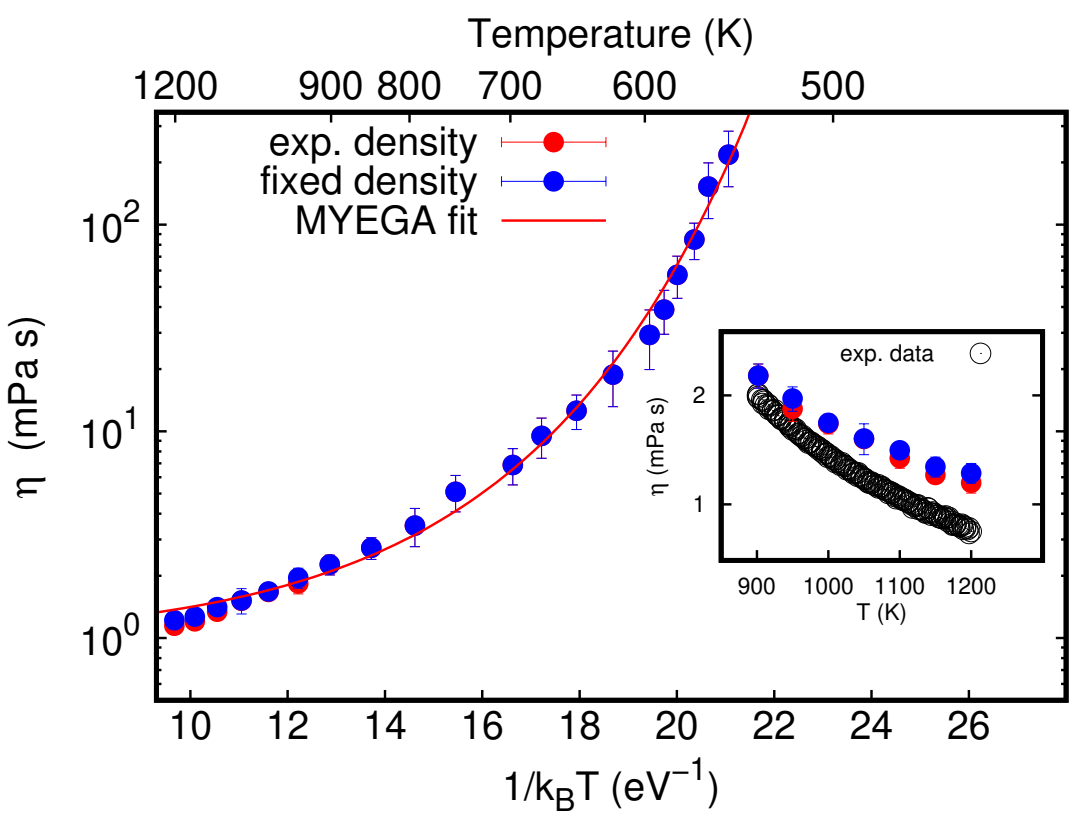

FIG. 6. Angell plot for the viscosity $\eta$ as a function of inverse temperature. The blue dots refer to simulations with the density fixed at the experimental density at $T_m$ (see Sec. II) while the red dots refer to simulations at the experimental density for $T > T_m$ from Ref.[22]. The error bars on $\eta$ are estimated from the block averages (see Sec. II). As a further quantification of the uncertainties in the viscosity, see Figure S10 in the supplementary material for the dependence of the results on the maximum time $t_{max}$ used in the evaluation of $\eta$ (see Sec. II). The continuous line is the fitting with a MYEGA function (see text). The comparison with the experimental data at high temperature from Ref.[22] is shown in the inset.

It turns out that the Maxwell relation between the viscosity and the $\alpha$-relaxation time ($\eta = G_\infty \tau_\alpha$) holds reasonably well in the whole temperature range with $G_\infty$= 2.9 GPa and with a larger deviation only at the two lowest tempera-tures investigated here, as shown in Figure 7. The relaxation time of the intermediate scattering function is used here as a good proxy for the relaxation time of the time-dependent shear modulus.[58] We mention that previous DFT simulations (PBE+D2)[24] yielded a slightly larger $G_\infty$ equal to 3.4 GPa in the temperature range 820-1800 K from the calculation of $D$ and the application of SER and from the ISF calculation of $\tau_\alpha$. The theoretical temperature dependent $G_\infty(T)$= $\eta(T)/\tau_\alpha(T)$ is shown in the inset of Figure 7a. Experimentally, a mild temperature dependence of $G_\infty(T)$ ranging from 2.8 to 3.1 GPa was observed by decreasing temperature from 1200 K to 930 K.[25] $G_\infty(T)$ was obtained from the coherent relaxation times measured by quasi-elastic neutron scattering.[25] The experimental temperature dependent $G_\infty(T)$ is then compared in Figure 7b with the theoretical ratio $G_\infty(T) = \eta(T)/\tau_\alpha^{cohe}(T)$ obtained from the coherent relaxation times $\tau_\alpha^{cohe}$. The Maxwell relation was shown to hold also in GeTe from NN-MD simulations in the range 520-1000 K ($T_m$=990 K)[33] and from experimental data in the range 850-1150 K.[25]

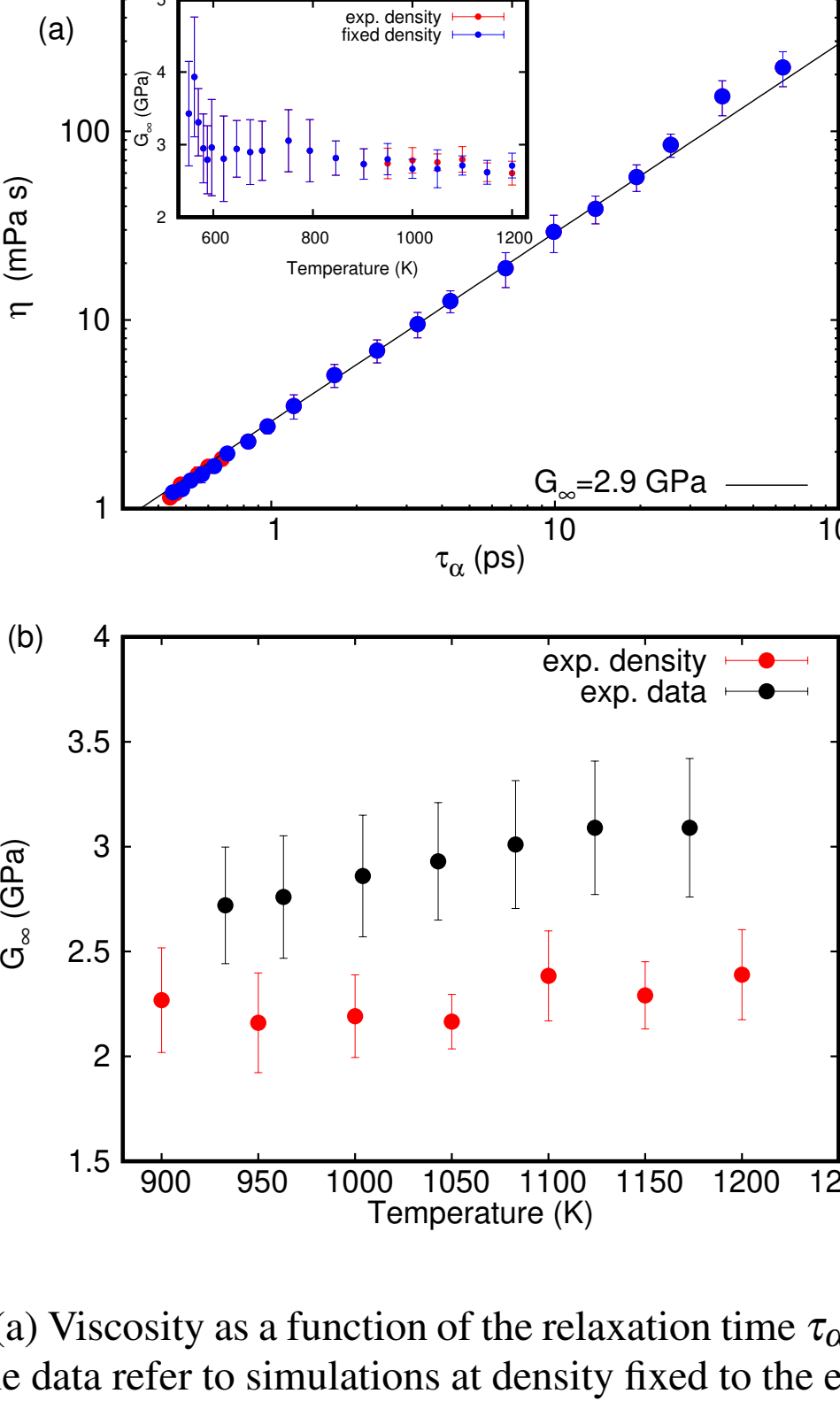

FIG. 7. (a) Viscosity as a function of the relaxation time $\tau_\alpha$ (from the ISF). The data refer to simulations at density fixed to the experimental density at $T_m$ (blue dots) and at the experimental temperature-dependent densities above $T_m$ (red dots). The linear fit of the Maxwell relation $\eta = G_\infty \tau_\alpha$ (continuous line) yields $G_\infty$= 2.9 GPa for both sets of simulations. The temperature dependent $G_\infty(T)$= $\eta(T)/\tau_\alpha(T)$ (blue dots) is shown in the inset. (b)The theoretical temperature dependent $G_\infty(T)$= $\eta(T)/\tau_\alpha^{cohe}(T)$ obtained from the coherent relaxation times $\tau_\alpha^{cohe}$ are compared with experimental data from Ref.[25] (black dots).

Coming back to the temperature dependence of $\eta$, we fitted our results with the Mauro-Yue-Ellison-Gupta-Allan (MYEGA) function[61]

$$\log_{10}\eta(T) = \log_{10}\eta_\infty + (12 - \log_{10}\eta_\infty)\cdot\frac{T_g}{T} \cdots \exp\left[\left(\frac{m}{12-\log_{10}\eta_\infty} - 1\right)\left(\frac{T_g}{T} - 1\right)\right], \tag{6}$$

with $\eta$ in Pa s. The MYEGA function contains explicitly as fitting parameters the glass transition temperature $T_g$ ($\eta(T_g)$=$10^{12}$ Pa s), the fragility index $m$, and the extrapolation of the viscosity at infinite temperature $\eta_\infty$. The best fit MYEGA function shown in Figure 6 yields $T_g$=391 K, $m$=94 and $\log_{10}\eta_\infty$=-3.02, or $T_g$=392 K, $m$=93 and $\log_{10}\eta_\infty$=-2.99 by fitting the data at fixed volume (blue dots in Figure 6). Our results are remarkably close to the DSC values of $T_g$=383 K, and $m \simeq$90 from Ref.[16]. This comparison needs, however, some comments. In the analysis of Ref.[16], the kinetic prefactor ($u_{kin}$) of the crystal growth velocities was inferred from the DSC traces for $T$ < 650 K. $u_{kin}$ was assumed to be proportional to the diffusion coefficient and then the viscosity was related to $u_{kin}$ via a fractional Stokes-Einstein relation $u_{kin} \propto \eta^{-\xi}$. The proportionality constant and the exponent $\xi$ were obtained in turn by setting $\eta$ at $T_m$ to the value estimated from ab-initio MD simulations[23] and $\eta(T_g) = 10^{12}$ Pa s, where $T_g$ was identified with the crystallization temperature. This fitting yields $\xi = 0.64$. As we will see in the next section, the exponent $\xi$ that best fit our data is $\xi$=0.84. Therefore, our data are only partially consistent to those of Ref.[16]. Moreover, as we have already mentioned, a recent DSC work[7] has shown that crystallization takes place in the amorphous phase for low heating rates, and only high heating rates allow identifying a $T_g$ of about 473 K from the endothermic peak. To reconcile our results with a $T_g$ of about 473 K, we must rely on some deviation with a sharp bend from a single MYEGA function for temperatures below 550 K. However, our results are unambiguously inconsistent with the fragile-to-strong crossover at the high temperature of $T_{fs}$= 683 K proposed in Ref.[20]. By inspection of the Angell plot in Figure 6, our data between 600 and 550 K might look on a straight line, i.e. displaying a strong Arrhenius behavior which mirrors the behavior of the diffusion coefficient in Fig. 1b. and the relaxation time in Fig. 4b. This temperature range is, however, too narrow, and the extrapolation of the Arrhenius function that fits the data in the 600-550 K range would assign an unphysical $T_g$ below 300 K (from $\eta(T_g)$=$10^{12}$ Pa s). Still, if we accept a strong behavior in the region 600-550 K, we must assume that another crossover with a larger slope would occur at lower temperature. As mentioned above, a deviation with a sharp bend from a single MYEGA function is needed to reconcile our data with the experimental $T_g$ of 473 K. We remark that our MYEGA fitting is indistinguishable from the fitting by a generalized MYEGA[62] with a double exponential expression and five parameters, often used to reproduce a FSC. For the generalized MYEGA form, only a single exponential term contributes significantly to our fitting.

We might wonder whether our calculations underestimate $\eta$ at low temperature because the system goes out of equilibrium by approaching $T_g$. We note that we computed $\eta$ by quenching from 900 K in 150 ps and then equilibrating the system at the target temperatures for 6.6-13.2 ns. We check this issue by re-computing $\eta$ starting from the system equilibrated at a higher temperature of 625 K for 4.4 ns and then cooling and further equilibrating the model for 4 ns at 590 K. The calculation yields $\eta$= 40 mPa s to be compared with the value of 39 mPa s obtained previously. By further cooling the same model from 590 to 580 K and equilibrating for 5 ns, we obtain $\eta$= 48 mPa s which is close to the value of 57 mPa s obtained previously at 580 K. Still, this does not prove that the system is in equilibrium. Indeed, an indication that the system is out of equilibrium at the lowest temperatures investigated here comes from the calculation of the specific heat $C_v$ shown in Figure 8. This is obtained from NVE simulations by cooling the liquid (blue dots in Figure 8) or heating the amorphous phase (red dots in Figure 8 ) previously generated by quenching from 1000 K to 260 K in 100 ps. The specific heat $C_v$ per particle and at constant volume was obtained from the temperature fluctuations in NVE simulations as[63]

$$C_v = \frac{3}{2}k_B(1-\frac{3}{2}N\frac{\Delta T^2}{T^2})^{-1} \quad (7)$$

$C_v$ is expected to feature a peak at $T_g$ on heating the amorphous phase and a downward jump on cooling the liquid.[58] Since the amorphous phase was generated by fast cooling, the peak on heating overestimates $T_g$. However, also the jump in $C_v$ on cooling is at a temperature at which our calculated viscosity is much lower than the value expected at $T_g$ ($10^{12}$ Pa s), which indicates that the system is out of equilibrium. Unfortunately, it is not possible to equilibrate the system in the temperature range 550-650 K for overall more than about 15 ns because of the onset of crystallization. In the real system as well and in particular in the devices, crystallization takes place in the same temperature range. Therefore, it is conceivable that the viscosity could neither be measured experimentally on a better equilibrated system in this temperature range. At even lower temperatures, the nucleation rate drops and then a better equilibration on longer times is possible, but on the other hand the relaxation time also becomes very long on a time scale which is accessible in experiments, but not by MD simulations. The energy as a function of temperature in NVT runs of cooling the liquid and heating the amorphous phase is shown in Fig. 9.

### D. Breakdown of Stokes-Einstein relation

In Figure 10(a) we show the ratio $\eta D/T$ that should be a temperature-independent constant whenever the SER holds. The viscosity is computed from NVT simulations and $D$ is computed in NVE simulations at a very similar average temperature (in most cases within 3 K, but for a larger misfit of 6 K at 700 K and of 4 K at 625 K). If both the Maxwell and the SER relations hold also the ratio $\tau_\alpha D/T$ should be a temperature-independent constant. This quantity is shown in Figure 10(b). But for small fluctuations due to the several uncertainties (see above), the SER reasonably holds down to 680 K. At lower temperatures, a breakdown of the SER clearly occurs with a steep increase in both the $\eta D/T$ and $\tau_\alpha D/T$ ratios.

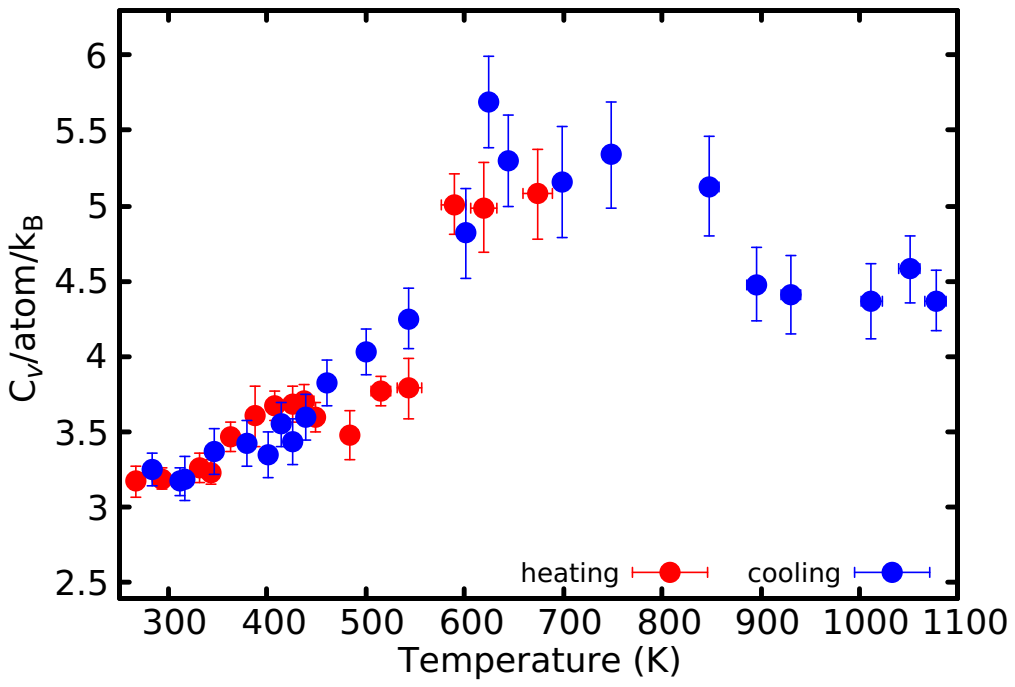

FIG. 8. Specific heat ($C_v$) per particle from NVE simulations on cooling (blue dots) from 3996-atom simulations, and on heating (red dots) a 999-atom amorphous model (with vdW, from Ref.[36]). The density is fixed to $\rho$=0.03075 atom/Å$^3$.

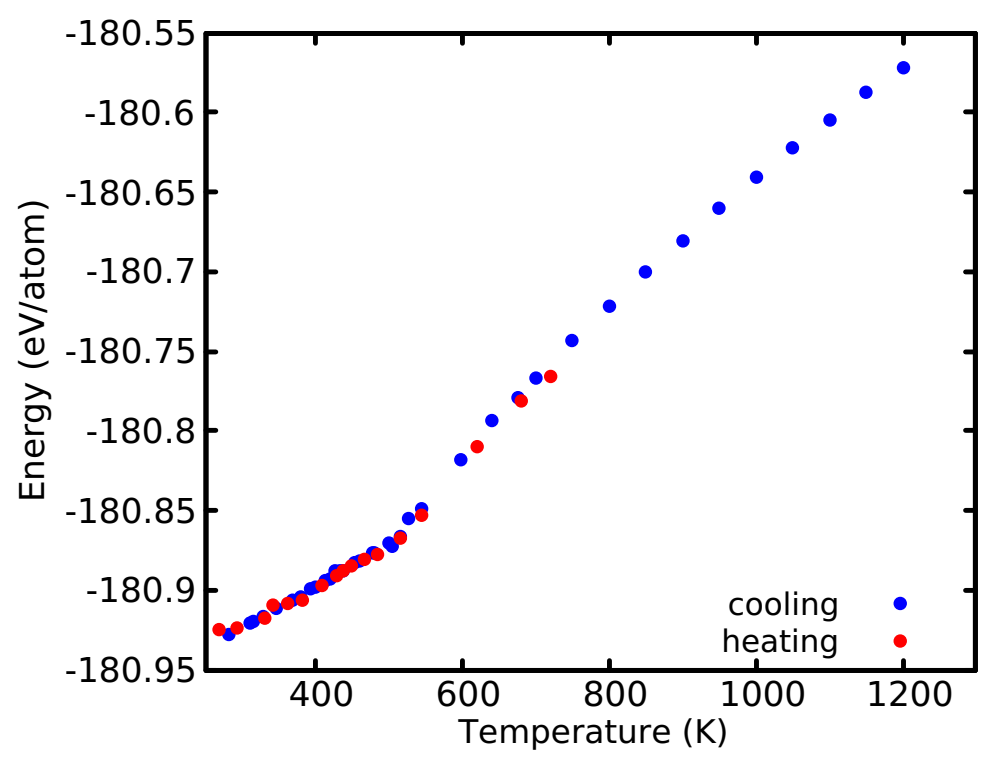

FIG. 9. Energy versus temperature in the two NVV runs of cooling the liquid (blue dots) and heating the amorphous phase (red dots).

A breakdown of SER with a rapid increase $\eta D/T$ by decreasing the temperature below 700 K was also reported for GeTe from NN-MD simulations.[32] Experimentally, an increase of the ratio $\eta D/T$ by a factor of about 1.3 was found in GST by decreasing the temperature from 1120 to 903 K.[25] In the same temperature range we observe some oscillations in the $\eta D/T$ ratio with a mild increase by decreasing temperature. The increase by about 1.3 seen experimentally seems outside our error bars. No direct experimental data on the viscosity or relaxation time are available in the supercooled liquid. To quantify the deviation from the SER, we use the fractional Stokes-Einstein relation[64]

$$D \propto (\tau_\alpha/T)^{-\xi} \;\; ; \;\; D \propto (\eta/T)^{-\xi} \quad (8)$$

with a fractional exponent $0 < \xi < 1$ ($\xi$=1 for SER). The fractional exponent can be found from a linear regression of the data $\log D$ vs $\log(\eta/T)$ (or $\log(\tau_\alpha/T)$) shown in Figure 11 that yields $\xi \simeq 0.84$ for $D \propto (\tau_\alpha/T)^{-\xi}$ and $\xi \simeq 0.82$ for $D \propto (\eta/T)^{-\xi}$. As already mentioned, our value of $\xi$ is higher than the value of 0.67 extracted from the analysis of the crystal growth velocity measured by DSC below $T_m$.[16] Above $T_m$ our $\xi$ is 1 (no breakdown) while experimentally is 0.63.[25] Overall the breakdown of SER in our simulations is less striking than in experiments.

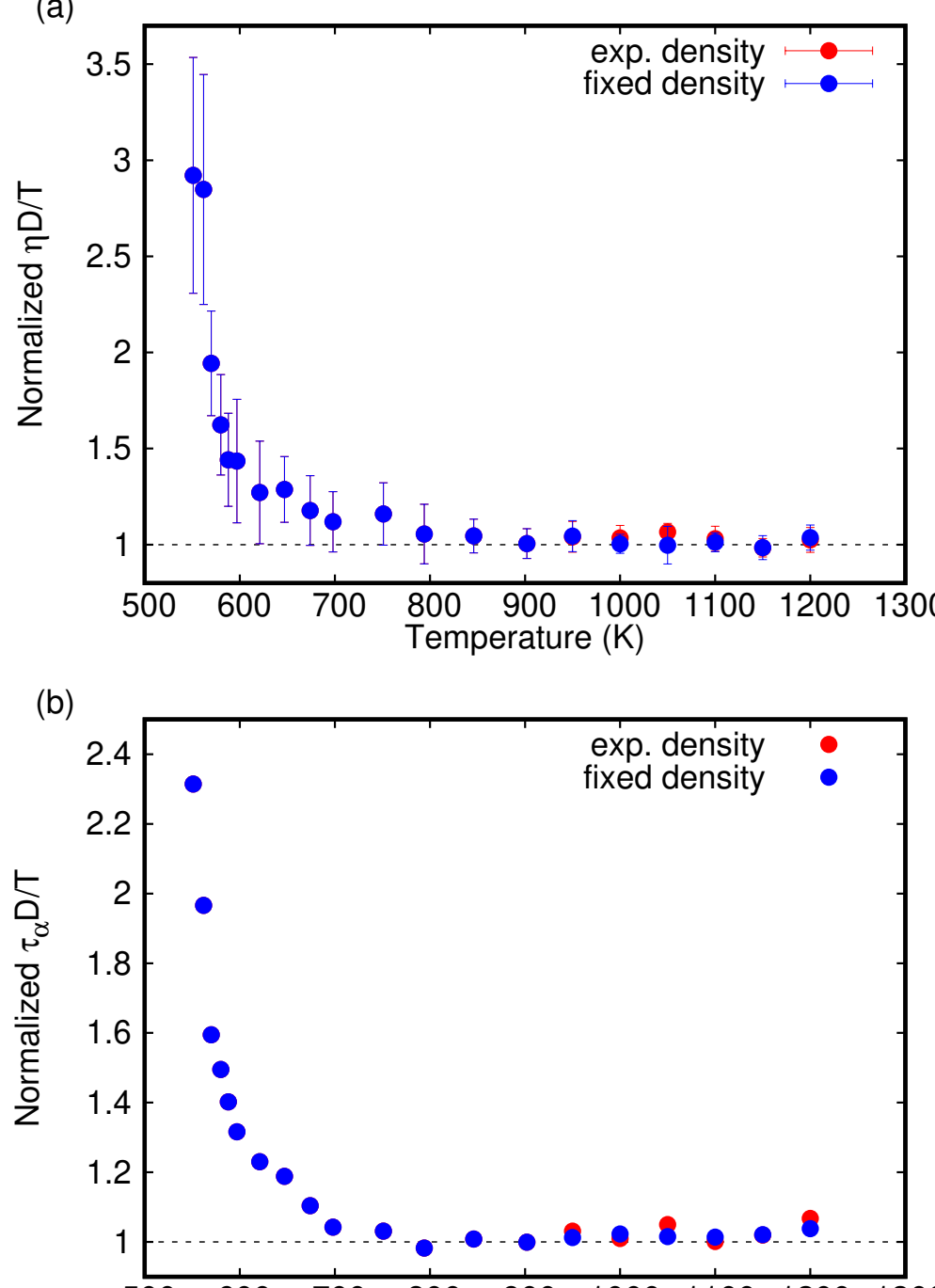

FIG. 10. The ratio (a) $\eta D/T$ and (b) $\tau_\alpha D/T$ as a function of temperature, normalized to their value at 900 K. When the SER holds, $\eta D/T$ is a temperature-independent constant. If both the SER and the Maxwell relation hold (see Eq. 7), $\tau_\alpha D/T$ should be a temperature-independent constant as well. A clear breakdown of SER occurs below 680 K. The data refer to simulations at density fixed to the experimental density at $T_m$ (blue dots) and at the experimental temperature-dependent densities above $T_m$ (red dots).

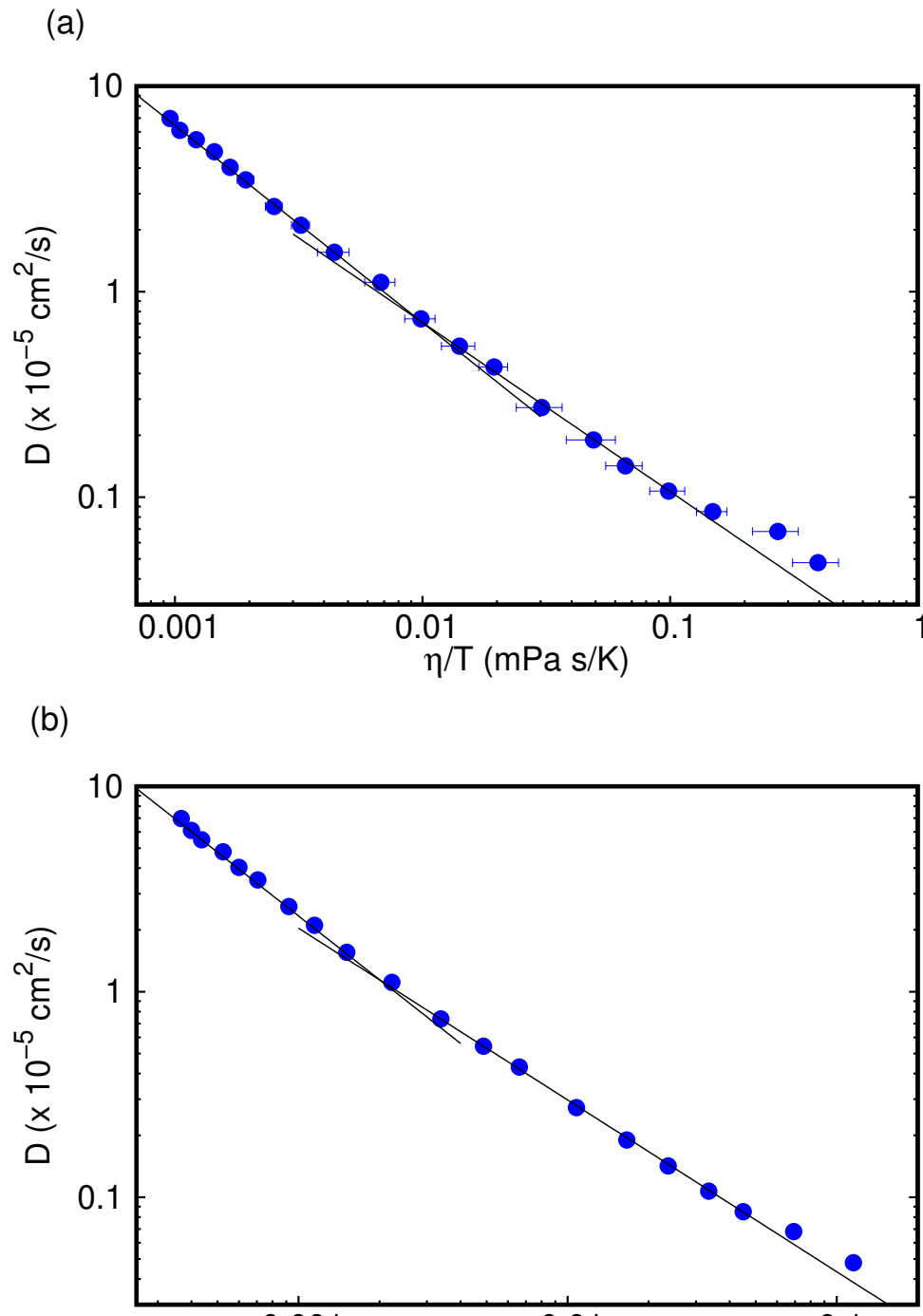

FIG. 11. Plot of $\log D$ versus (a) $\log(\eta/T)$ or (b) $\log(\tau_\alpha/T)$. The kink indicates a deviation from the SER that is quantified by the fractional exponent $\xi$ (see Eq. 8). Linear regression yields $\xi$=1 at high temperature when the SER holds and $\xi \simeq 0.83$ below 680 K.

The decoupling between viscosity and atomic mobility responsible for the breakdown of SER can be ascribed to a dynamical heterogeneity (DH) in the supercooled liquid, which is another characteristic feature of fragile liquids.[58,65–69] A heterogeneous dynamics is likely to contribute also to the formation of a non-exponential (stretched) relaxation of the ISF discussed in Sec. IIIB. Dynamical heterogeneity arises from the formation of spatially localized regions in which atoms move substantially faster than the average and regions where atoms move slower than the average. The presence of DH leads to the breakdown of SER because, roughly speaking, $D$ is controlled by the fast moving atoms, while structural relaxations ($\tau_\alpha$ and $\eta$) are controlled by the regions with slow moving atoms. The argument is the following: if we assume that the SER holds locally in regions with different mobility, the average $< D >$ can be obtained by applying locally the SER, i.e. $< D > /T \propto < \eta^{-1} >$ where, however, $< \eta^{-1} >$ differs from $< \eta >^{-1}$ when the average is over a broad distribution.[21,58] Dynamical heterogeneity can be studied by MD as well,[67–69] as we already did for the parent compound GeTe in a previous work.[34] A similar analysis for GST is discussed in the next section.

### E. Dynamical heterogeneity

To unravel the emergence of DH in supercooled liquid we computed the dynamical susceptibility $\chi_4(q,t)$,[70,71] defined by the variance of the ISF (see Eq. 2) as:

$$\chi_4(q,t) = N\left[\left\langle |\Phi_s(q,t)|^2 \right\rangle - \langle \Phi_s(q,t) \rangle^2\right] \tag{9}$$

The dynamical susceptibility $\chi_4(q,t)$ quantifies the fluctuations in the correlation of single atom displacements over the time and length scales given by $t$ and $q$. We computed $\chi_4(q_o,t)$ for the same $q_o$= 2 Å$^{-1}$ used in the analysis of the ISF (main peak of the static structure factor, see Sec. IIIB) and for five different temperatures 900, 750, 650, 600, 550 K as shown in Figure 12. The function $\chi_4(q_0,t)$ starts from zero at short times, reaches a maximum between the $\beta$ and the $\alpha$-relaxation regimes, and then approaches one at long times. The height of the peak increases by decreasing temperature and is proportional to the number of atoms involved in cooperative motions in the spatial regions of slow and fast moving atoms.[72]

The time $t^*$ corresponding to the maximum of $\chi_4(q_o,t)$ gives a measure of the time over which we expect to see a cor-relation in the velocities of the atoms in the regions of most mobile (MM) and most immobile (MI) particles (that we will define in a quantitative manner later on). In other words, $t^*$ is a characteristic time over which an atom keeps its character of MM or MI particle. $t^*$ corresponds to 1.7, 3.3, 8.1, 20.4 and 62 ps at 900, 750, 650, 600, and 550 K, respectively.

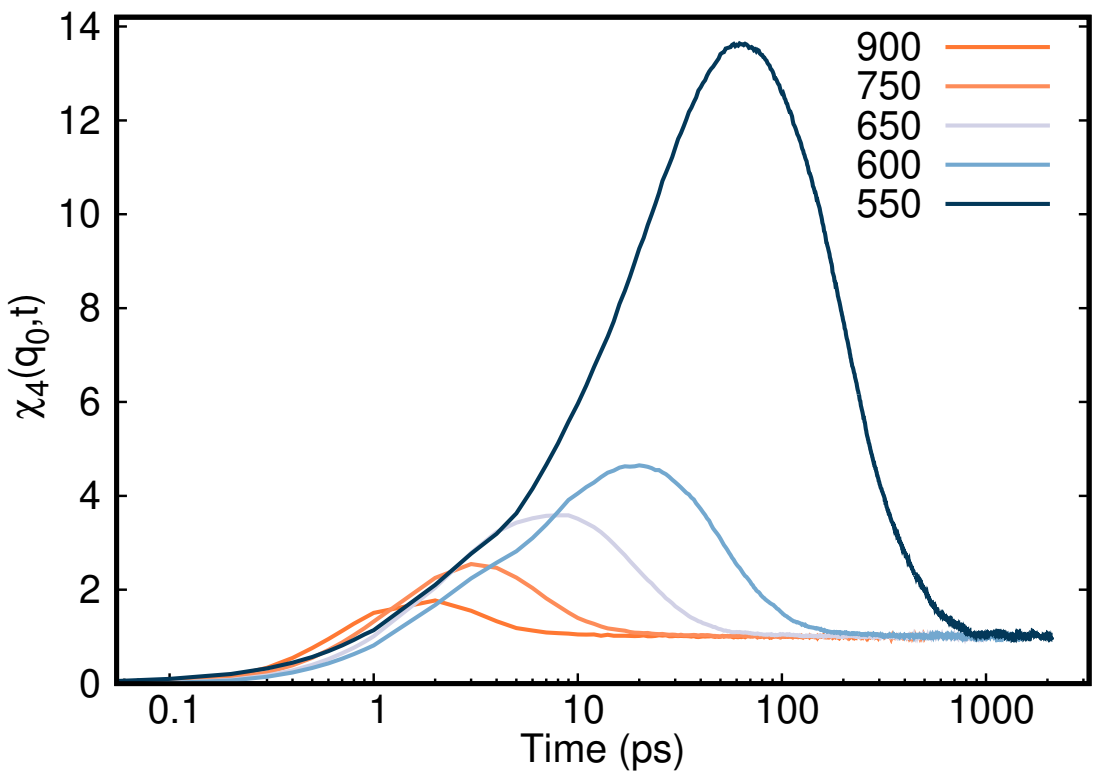

FIG. 12. The dynamical susceptibility $\chi_4(q_o,t)$ (Eq. 9) as a function of time for different temperatures and for $q_o$= 2 Å$^{-1}$ (see text).

We used our estimate of $t^*$ for the so-called iso-configurational analysis that can be used to obtain a spatially-resolved map of dynamical heterogeneities[73] and then some hints on the possible structural origin of the clustering of atoms in MM and MI regions. In the description of this analysis, we follow closely our previous work on the study of DH in GeTe by NN-MD.[34]

The analysis consists of performing several (50 in our case) MD simulations starting from the same atomic configuration extracted from an equilibrium MD trajectory of liquid, but randomly choosing the initial velocities for each run from a Maxwell-Boltzmann distribution at a fixed temperature. We then performed averages over this ensemble of trajectories to establish correlations between the MM and MI regions and structural features present in the starting configuration. Different iso-configurational ensembles (four in our case) are generated starting from different atomic configurations. This procedure is repeated for different temperatures. We quantify the tendency of each atom $i$ to move by computing the so called dynamical propensity (DP) over the characteristic time $t^*$ defined above:

$$DP_i(t^*) = \left\langle \frac{(\vec{r}_i(0) - \vec{r}_i(t^*))^2}{MSD_{\alpha_i}} \right\rangle_{ISO}, \qquad (10)$$

where $\alpha_i$ denotes the atomic species of atom $i$ (Ge, Sb or Te), $MSD_{\alpha_i}$ is the MSD of that species for the same time $t^*$, and $\langle \ldots \rangle_{ISO}$ stands for the average over the $N_{ISO}$ trajectories of the iso-configurational ensemble. The normalization factor $MSD_{\alpha_i}$ is introduced because different species have different mobilities: atoms whose tendency to move is close to the average for their species have $DP \approx 1$, whereas deviations from unity correspond to atoms more or less mobile than the average. The distribution of the DP obtained from a single iso-configurational ensemble at three different temperatures is shown in Figure 13. Atoms with a DP at least one standard deviation away from the mean of the DP distribution are defined as most immobile (MI) or most mobile (MM, see Figure 13). This choice ensures that MI and MM clusters contain a representative fraction of atoms (about 10%) at all temperatures. Atoms are defined to belong to the same cluster if they are within a distance of 3.2 Å. We observed a clear increase in the size of MM and MI clusters by decreasing temperature. The size of the largest MM and MI clusters at the different temperatures from the four iso-configurational ensemble are shown in Figure 14.

As we did in our previous work on GeTe,[34] to visualize the MM and MI clusters we considered a DP density

$$\rho(DP)_i = \frac{1}{N_j} \sum_{j, r_{ij} < r_d}^{N_j} DP_j, \qquad (11)$$

where the sum runs over the neighboring atoms $j$ of atom $i$ up to a certain cutoff distance $r_d$, which we set equal to 3.2 Å.

In Fig. 15 we show the starting configuration of one iso-configurational analysis at three different temperatures (900, 650, and 500 K), where each atom is colored according to its DP density. At 900 K slow (blue) and fast (red) atoms are randomly distributed, while by decreasing temperature slow and fast moving atoms tend to cluster in spatially separated domains.

In our previous work on GeTe,[34] we found that the MM clusters feature a significantly larger concentration of Ge-Ge homopolar bonds. The emerging picture was that DH originates from the chemical specificity of GeTe in which structural heterogeneities in the form of the chains of Ge-Ge bonds preserve and enhance the atomic mobility very close to $T_g$. On these premises, we first computed the abundance of the different types of bonds in the MM and MI largest clusters which, however, does not seem to change significantly with respect to the bulk average at all temperatures. As opposed to GeTe, dynamical heterogeneities do not seem to correlate with chemical heterogeneities. We therefore tried to correlate the MM and MI regions not with the types of atoms involved in the bonds, but with the bond geometry, namely with the size of the Peierls distortion consisting of the formation of long and short axial bond around Ge and Sb atoms. In fact, DFT-MD simulations in Ref.[22] provided a correlation between the enhancement of the Peierls distortion by decreasing the temperature in liquid GST above $T_m$ and a deepening of the pseudogap at the Fermi level in the electronic density of states. A similar analysis in the supercooled liquid below $T_m$ suggested a correlation between the enhancement of the Peierls distortion and the metal-semiconductor transition.[74] Therefore, the Peierls distortion seemed a reasonable feature to look at in the attempt to correlate the dynamical heterogeneity with a structural property. To this end, we analyzed the angular-limited three-body correlation function (ALTBC)[75] which highlights the presence of a short and a long axial bond in defective octahedral configurations. Namely, ALTBC measures the probability of finding a bond with length $r_1$ mostly aligned (within a threshold angle of $25^o$) with a second bond of length $r_2$ formed with the same central atom. The ALTBC function in the ($r_1$, $r_2$) plane is shown in Figure 16 for central Ge atom and in Figures S11 and S12 in the supplemental material for central Sb and Te atoms, at the three different temperatures, for the bulk and for the two largest MM and MI clusters. The calculations are averaged over different snapshots (from 2 to 20) from all trajectories (50) of the four iso-configurational ensembles. The main feature that emerges from the analysis of the ALTBC is the enhancement in the MM region at 550 K of the tails of the distribution for longer second bonds for Ge. The secondary maxima at 3.6 and 3.8 Å are at distances too long to correspond to a real weak axial bond. Therefore, this feature can not be read as an enhancement of the Peierls distortion. Nevertheless, the presence of a neighboring atom on an axial position just outside the first coordination shell might provide a favorable configuration for the migration of a Ge atom. These tails are instead less evident in the MI regions than in the bulk. Moreover, a secondary peak corresponding to a short bond below 2.8 Å and suggesting a better defined Peierls distortion is present for MI Ge atoms. A larger Peierls distortion implies a stronger short bond and then a higher barrier for diffusion which is consistent with its enhancement in MI regions. We note that Ge is the most mobile species and that the difference in mobility between Ge and Sb or Te increases by decreasing temperature (see Figure S2 in the supplementary material). Therefore, an enhancement of the mobility of Ge in the MM regions via the mechanism depicted above could drag also the Sb and Te atoms in the same regions. In the ALTBC of Sb in the MM regions (Figure S11 in the supporting material), we note instead an enhancement of the Peierls distortion at 550 K (the appearance of a double feature in the main peak) which seems at odds with the expectation that a larger Peierls distortion would correpond to a stronger short bond and then possibly to a lower mobility. However, for the reasons outlined above, the nature of MM or MI region is probably ruled by the most mobile species (that drag the others), i.e. by Ge.

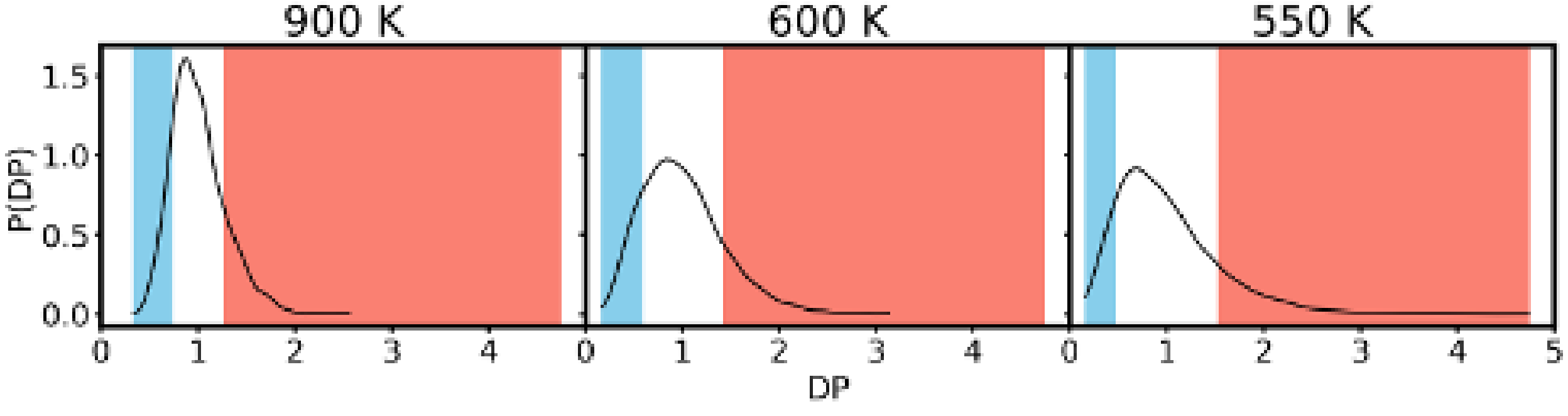

FIG. 13. Probability density distribution of DP at three different temperatures. Particles having a DP within blue or red regions are labeled as most immobile (MI) or most mobile (MM) particles respectively.

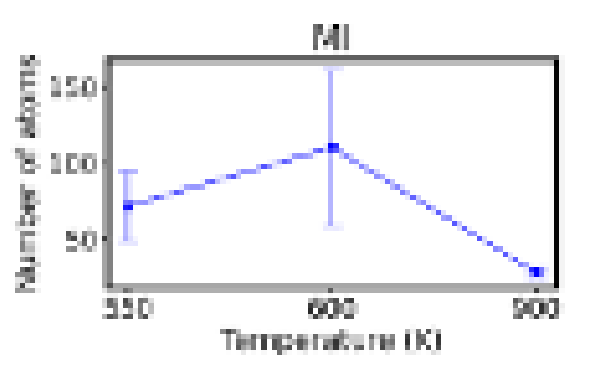

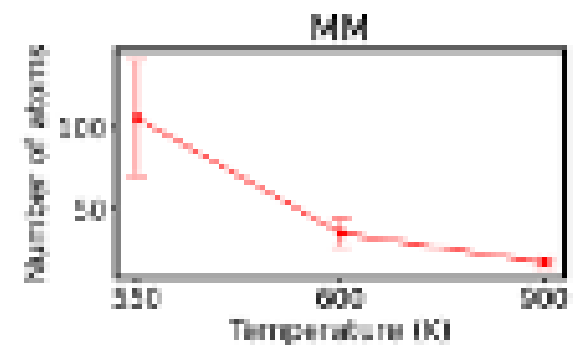

FIG. 14. Average size of the largest MI (left) and MM (right) clusters from four iso-configurational ensembles at each temperature. The error bar indicates the variance over the four iso-configurational ensembles.

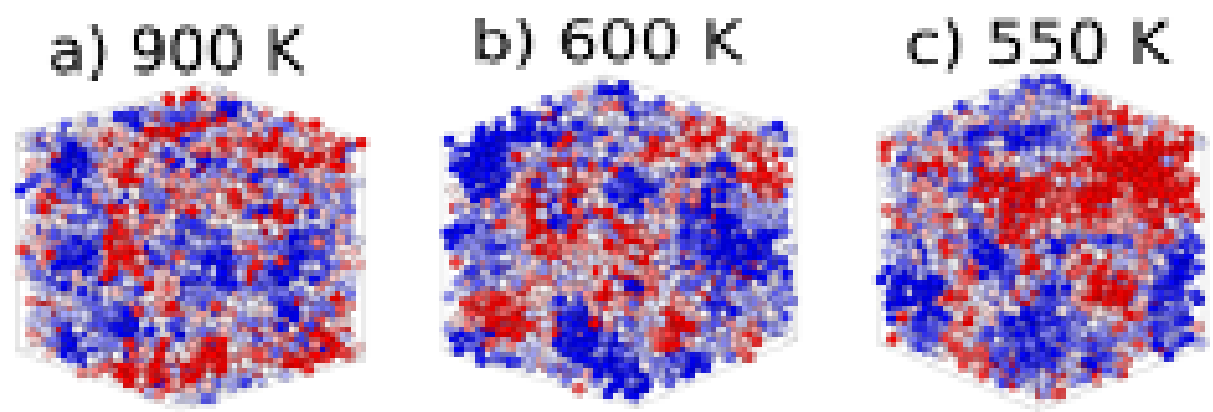

FIG. 15. Color map of the dynamical propensity density $\rho(DP)$ (see Eq. 11) for the starting configuration of one iso-configurational analysis at three temperatures (a) 900 K, (b) 600 K, and (c) 550 K. Atoms are colored according to their DP density, with MM atoms in red and MI atoms in blue.

## IV. CONCLUSIONS

In summary, by leveraging a MLIP that we recently developed, we performed MD simulations of the liquid and supercooled liquid phase of $Ge_2Sb_2Te_5$ to compute the viscosity $\eta$ from the Green-Kubo formula, the diffusion coefficient $D$ from the square displacement and the $\alpha$-relaxation time $\tau_\alpha$ from the incoherent and coherent intermediate scattering functions. The calculations have been performed down to 550 K, i.e. about 80 K above the latest experimental estimate of the glass transition temperature (470 K),[7] that required simulations lasting up to 18 ns. The fitting of $\eta$ with a MYEGA function provided a fragility index $m$=94 which is close to previous experimental estimate from DSC.[16] However, a deviation of $\eta$ from a single MYEGA function at lower $T$ is necessary to reconcile our calculated viscosity with the experimental $T_g$ of 470 K. Although we found some evidence of a fragile-to-strong crossover (FSC) in the range 590-550 K, the resulting activation energy in the strong regime is too low and implies an unphysical $T_g$ below 300 K. Therefore, a crossover at further lower temperatures into a regime with a higher fragility/activation energy is needed to meet the experimental $T_g$. The comparison of $\eta$, $D$ and $\tau_\alpha$ allowed us to assess the validity of the Maxwell relation down to the lowest temperature and the breakdown of the Stokes-Einstein relation (SER) below 680 K. We have also provided evidence of the presence of dynamical heterogeneities at the lowest temperatures responsible for the breakdown of SER. Iso-configurational analysis provided a tentative correlation between the dynamical heterogeneity and a particular structural feature of Ge atoms, namely the enhancement in the MM regions of the presence of an axial long contact (too long to be a weak bond) that could lower the activation barrier for diffusion and an enhancement of the Peierls distortion in the MI regions.

FIG. 16. ALTBC function for Ge atoms at (top panels) 900 K (central panels) 600 K and (lower panels) 550 K for the bulk (left panels), the two largest MI clusters (central panels) and the two largest MM clusters (right panels).

## SUPPLEMENTARY MATERIAL

See supplementary material for additional figures on the dynamical and structural properties of the supercooled liquid phase and for a synoptic table of all simulations performed.

## CODE AVAILABILITY

LAMMPS, and DeePMD are free and open source codes available at https://lammps.sandia.gov and https://deepmodeling.com/space/DeePMD-kit, respectively. The MLIP generated in O. Abou El Kheir et al., npj Comput. Mater. **10**, 33 (2024) is available in the Materials Cloud repository at https://doi.org/10.24435/materialscloud:a8-45.

## DATA AVAILABILITY STATEMENT

The data that support the findings of this study are available from the corresponding author upon reasonable request.

## AUTHOR CONTRIBUTIONS

M.B. conceptualized the work and wrote the original draft of the paper. The simulations were performed by S.M., O.A.E.K. and R.P.. All authors contributed to the analysis of the data and approved the final version of the paper.

## CONFLICTS OF INTEREST

There are no conflicts of interest to declare.

## ACKNOWLEDGMENTS

The project has received funding from European Union Next-Generation-EU through the Italian Ministry of University and Research under PNRR M4C2I1.4 ICSC Centro

Nazionale di Ricerca in High Performance Computing, Big Data and Quantum Computing (Grant No. CN00000013). We acknowledge the CINECA award under the ISCRA initiative, for the availability of high-performance computing resources and support.

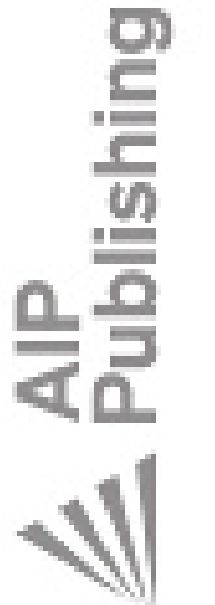

# Viscosity, breakdown of Stokes-Einstein relation and dynamical heterogeneity in supercooled liquid $Ge_2Sb_2Te_5$ from simulations with a neural network potential

## Supplementary Material

Simome Marcorini,[1] Rocco Pomodoro,[1] Omar Abou El Kheir,[1] and Marco Bernasconi[1]
[1]*Department of Materials Science, University of Milano-Bicocca, Via R. Cozzi 55, I-20125 Milano, Italy*

(*Electronic mail: marco-bernasconi@unimib.it)

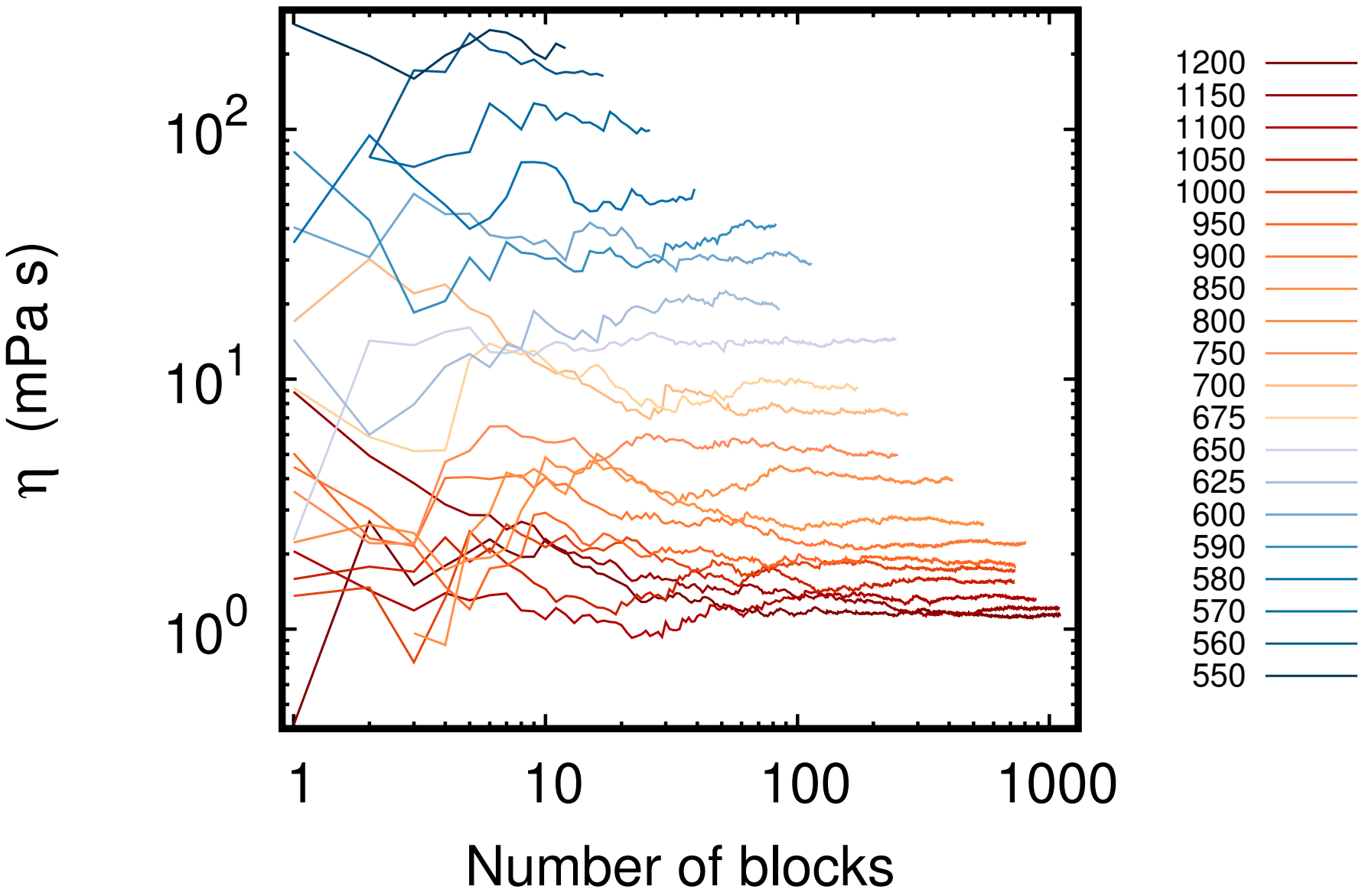

Figure S1. Convergence of the viscosity $\eta(t_{max})$ at time $t_{max}$ as a function of the number of blocks ($2t_{max}$ long) used in the block average (see Section II in the article).

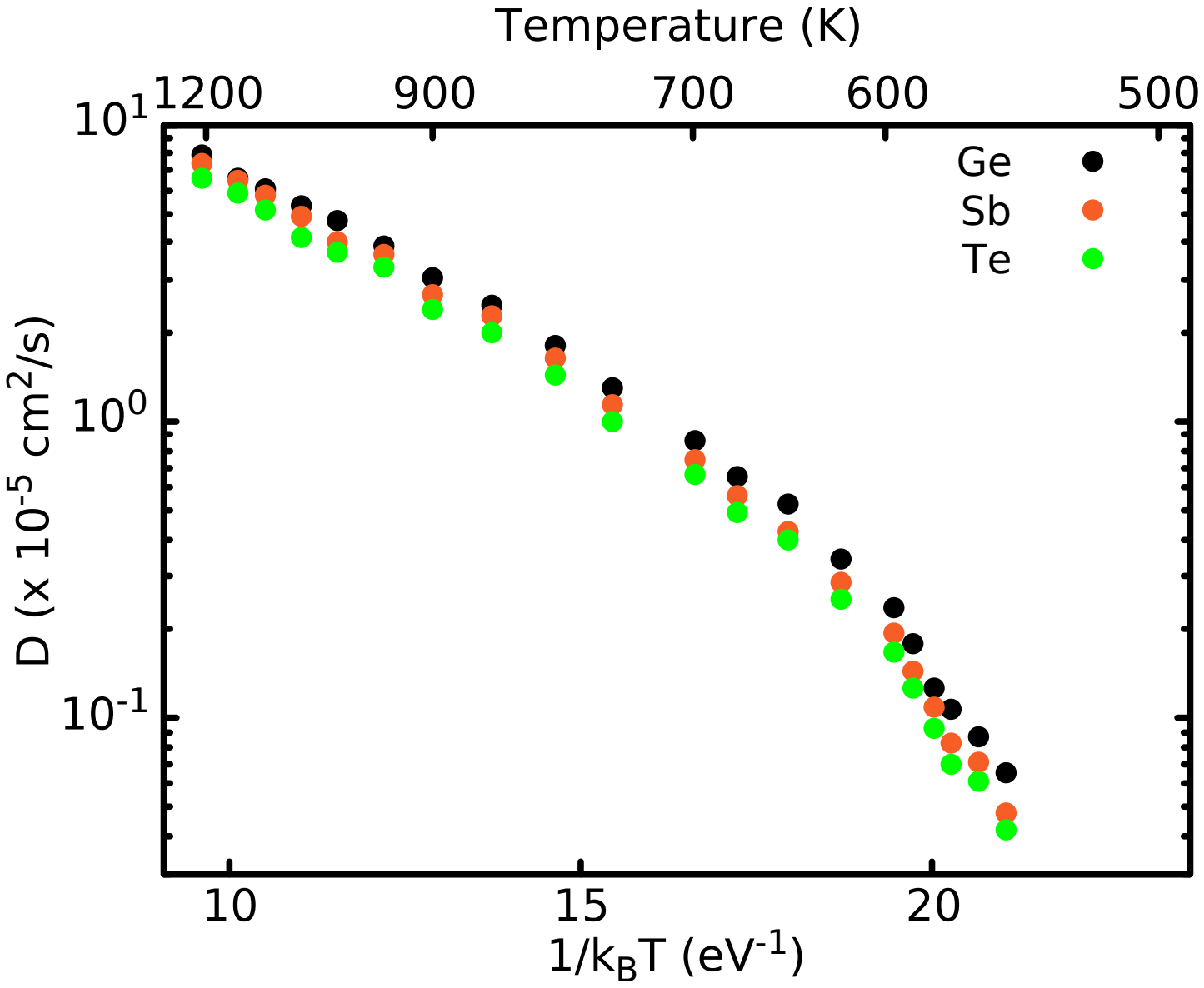

Figure S2. Diffusion coefficient as a function of temperature resolved for the different species.

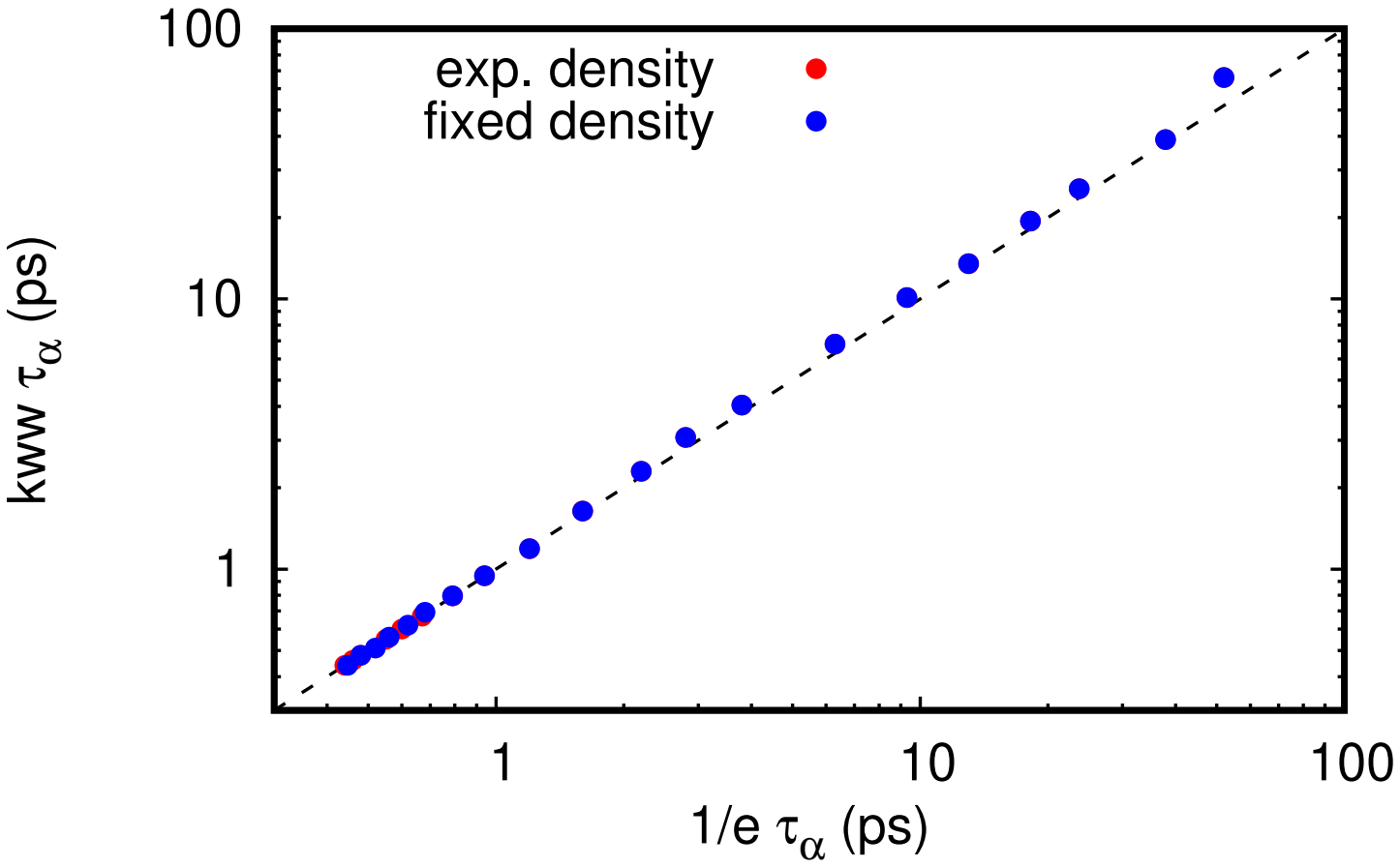

Figure S3. Comparison of $\tau_\alpha$ obtained from the KWW fit at long times and from the condition $F_s(q_o,t) = 1/e$. The data refer to different temperatures.

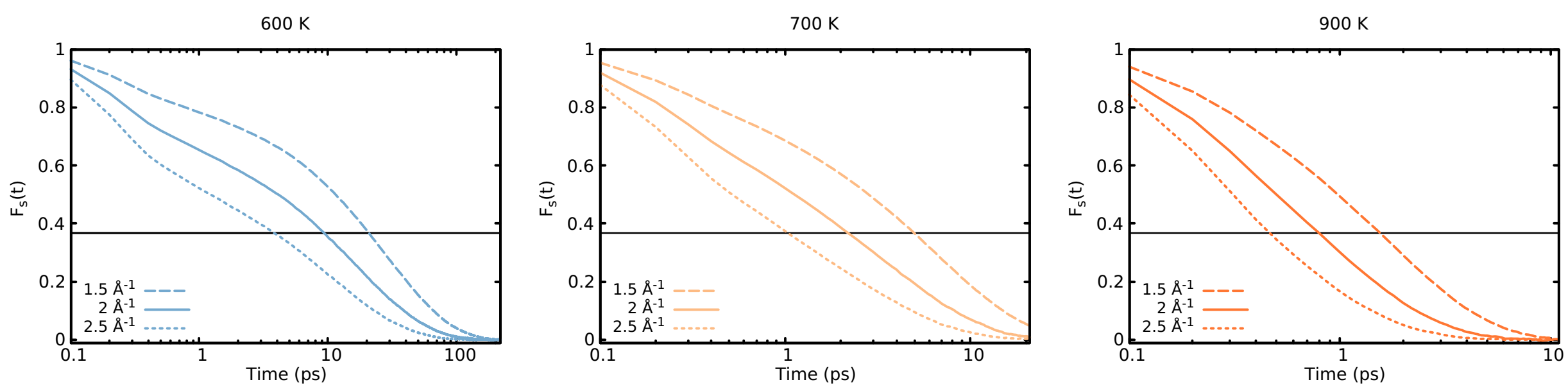

Figure S4. Incoherent intermediate scattering function $F_s(q,t)$ for three different values of $q$ and at the three different temperatures.

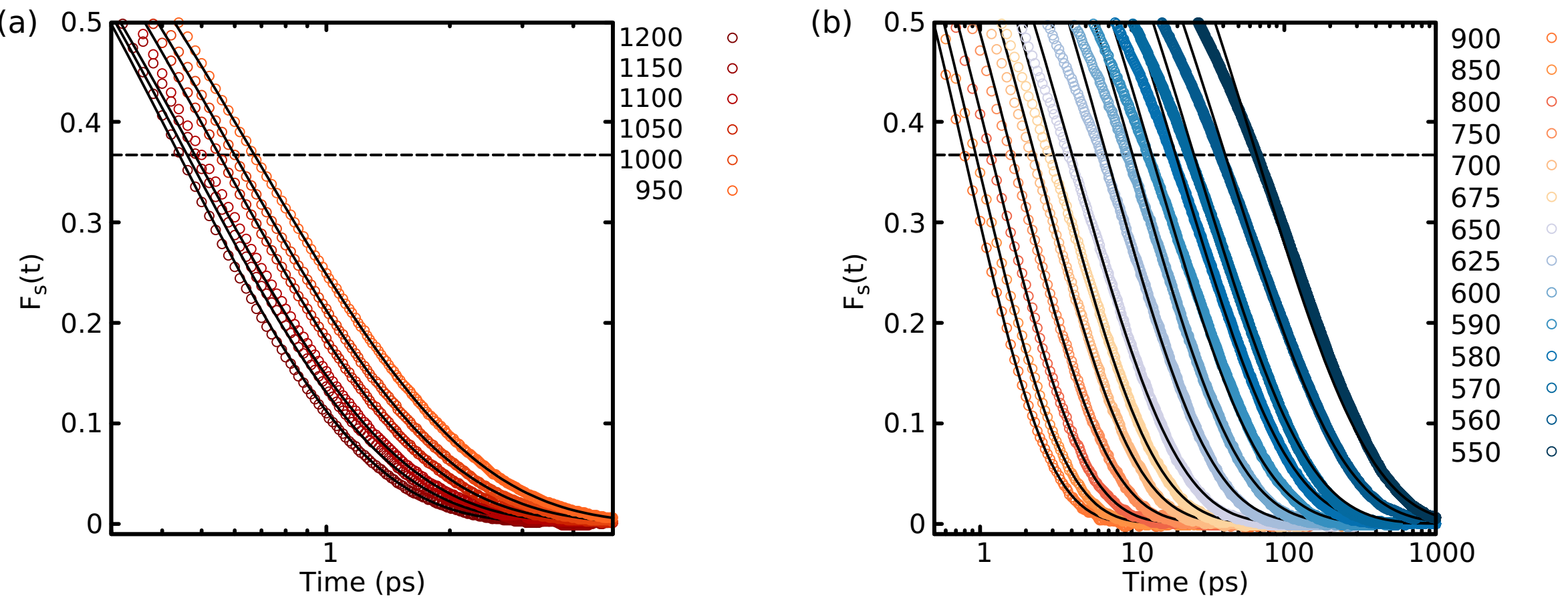

Figure S5. KWW fit (continuous lines) of the ISF (open dots, see Figure 2 in the article) for temperatures (a) above $T_m$ and (b) below $T_m$.

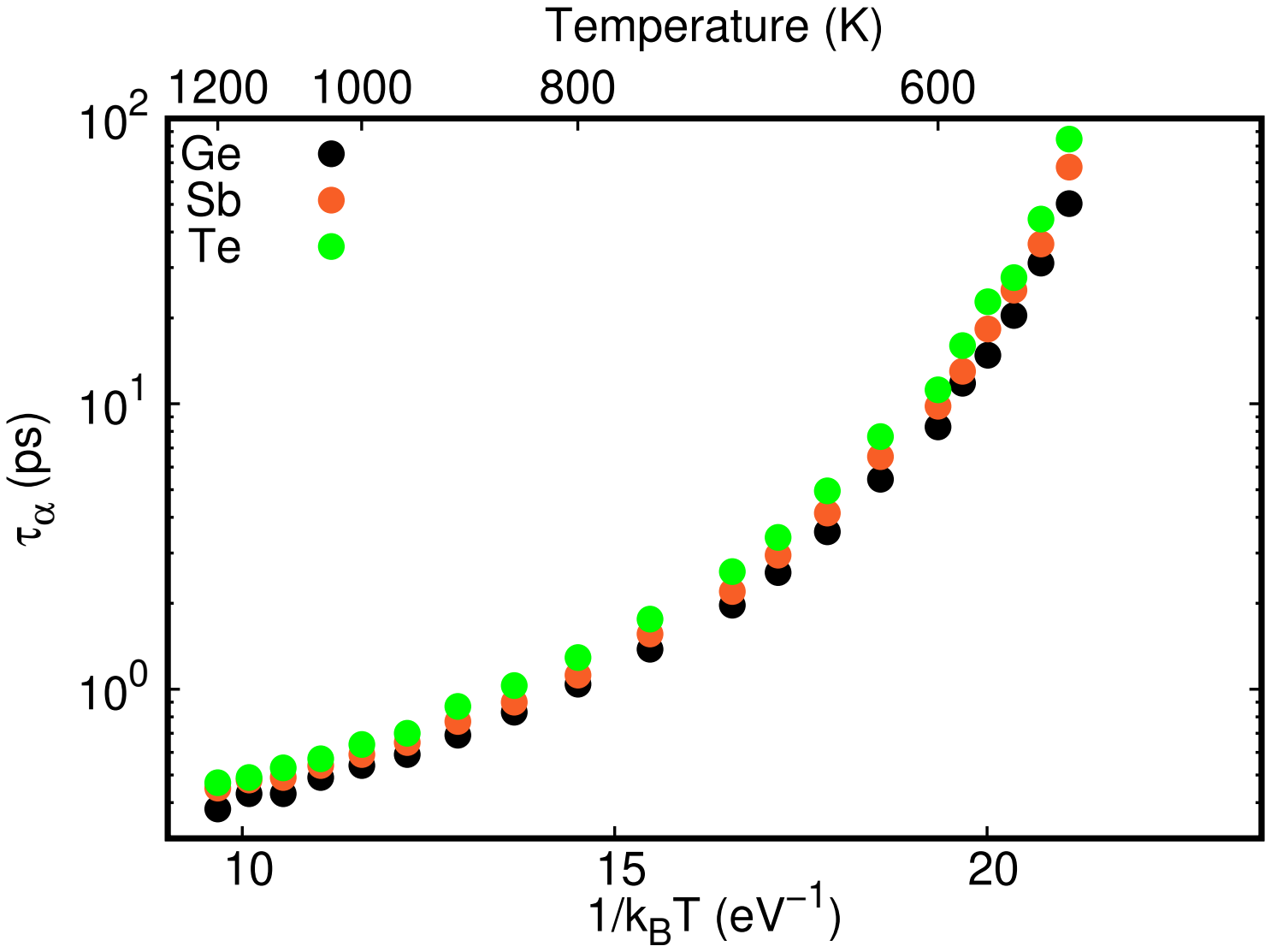

Figure S6. The relaxation time $\tau_\alpha$ as a function of temperature resolved for the different species. For the data above $T_m$ we used the values at the experimental density.

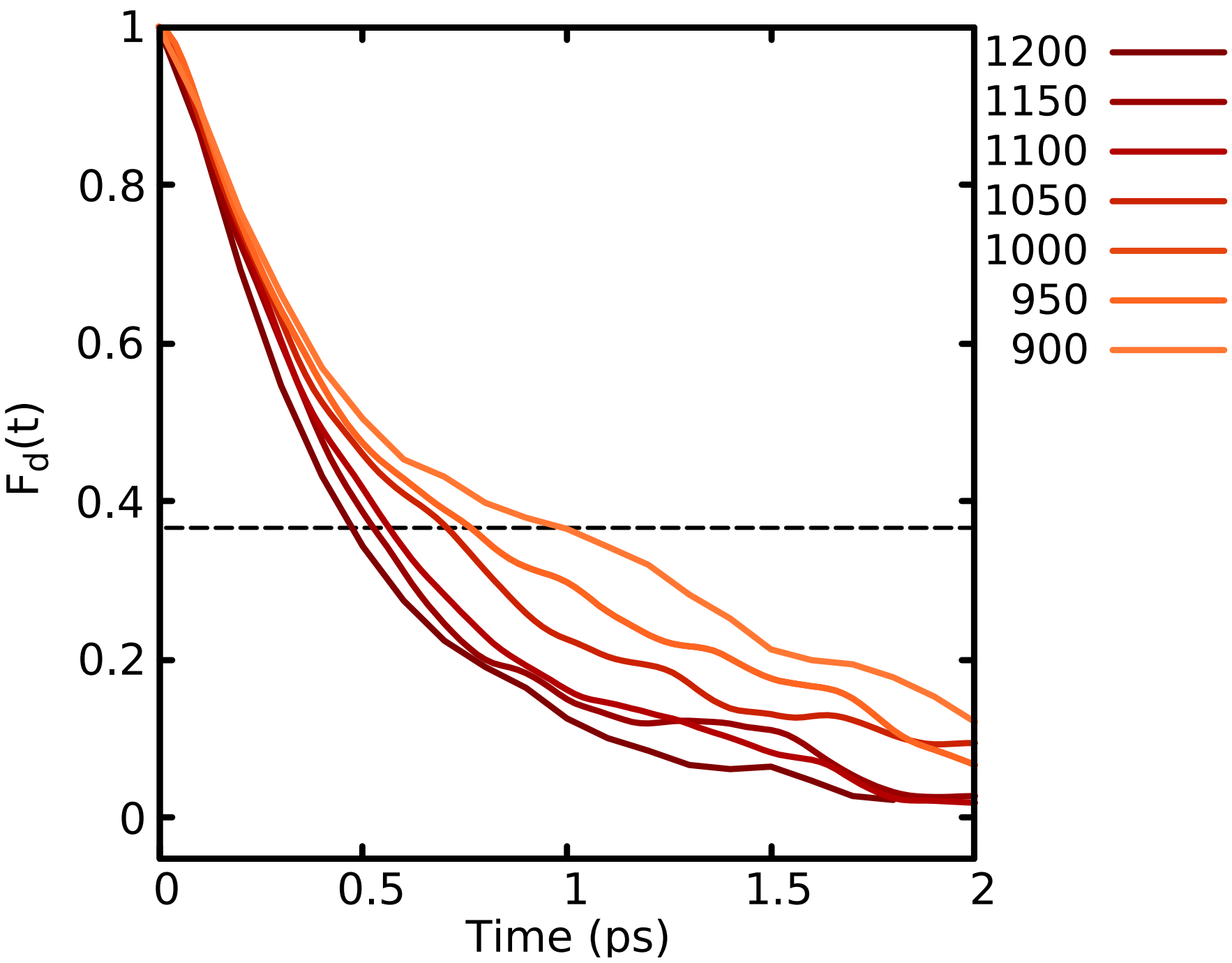

Figure S7. Coherent (distinct) intermediate scattering function $F_d(q_o,t)$ as a function of time at different temperatures above $T_m$ from NVT simulations. We chose $q_o$ = 2 Å$^{-1}$ that corresponds to the main peak in the experimental static structure factor (see article). The horizontal line indicates the condition $F_d(q_o,t) = 1/e$.

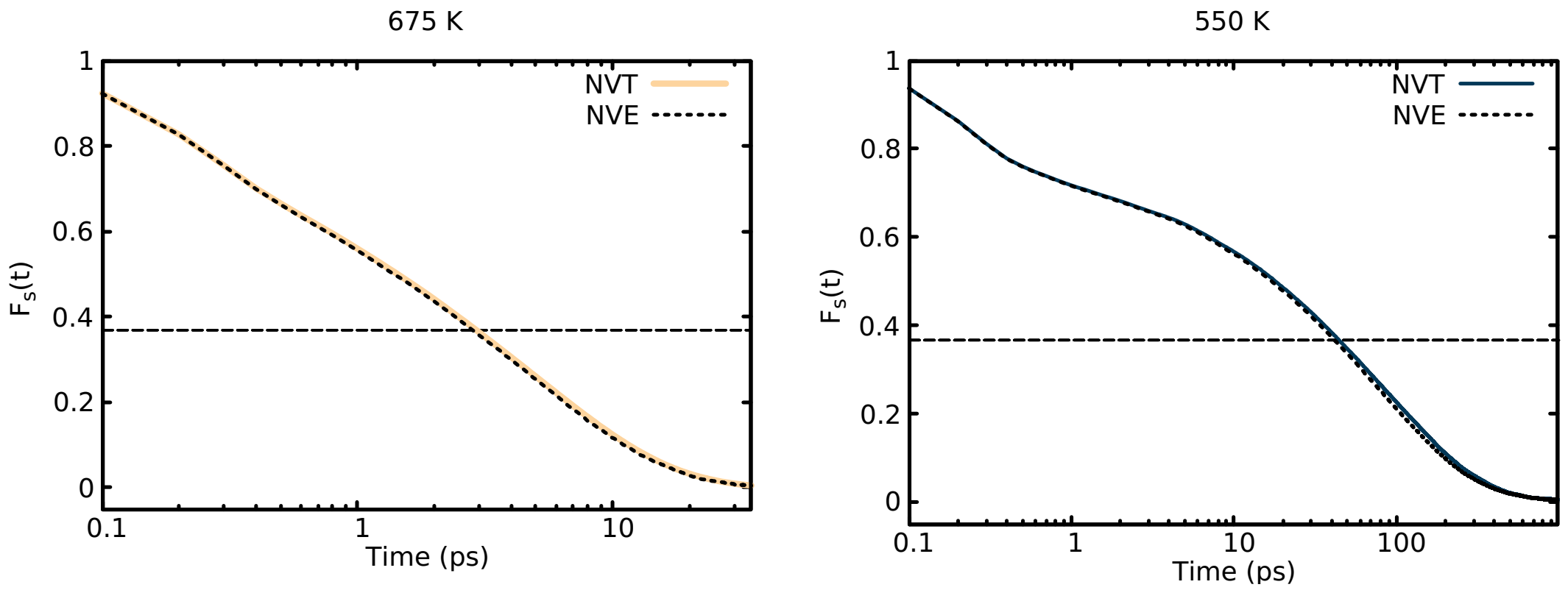

Figure S8. Incoherent intermediate scattering functions $F_s(q_o,t)$ obtained from NVT and NVE simulations, of the same length, at two different temperatures.

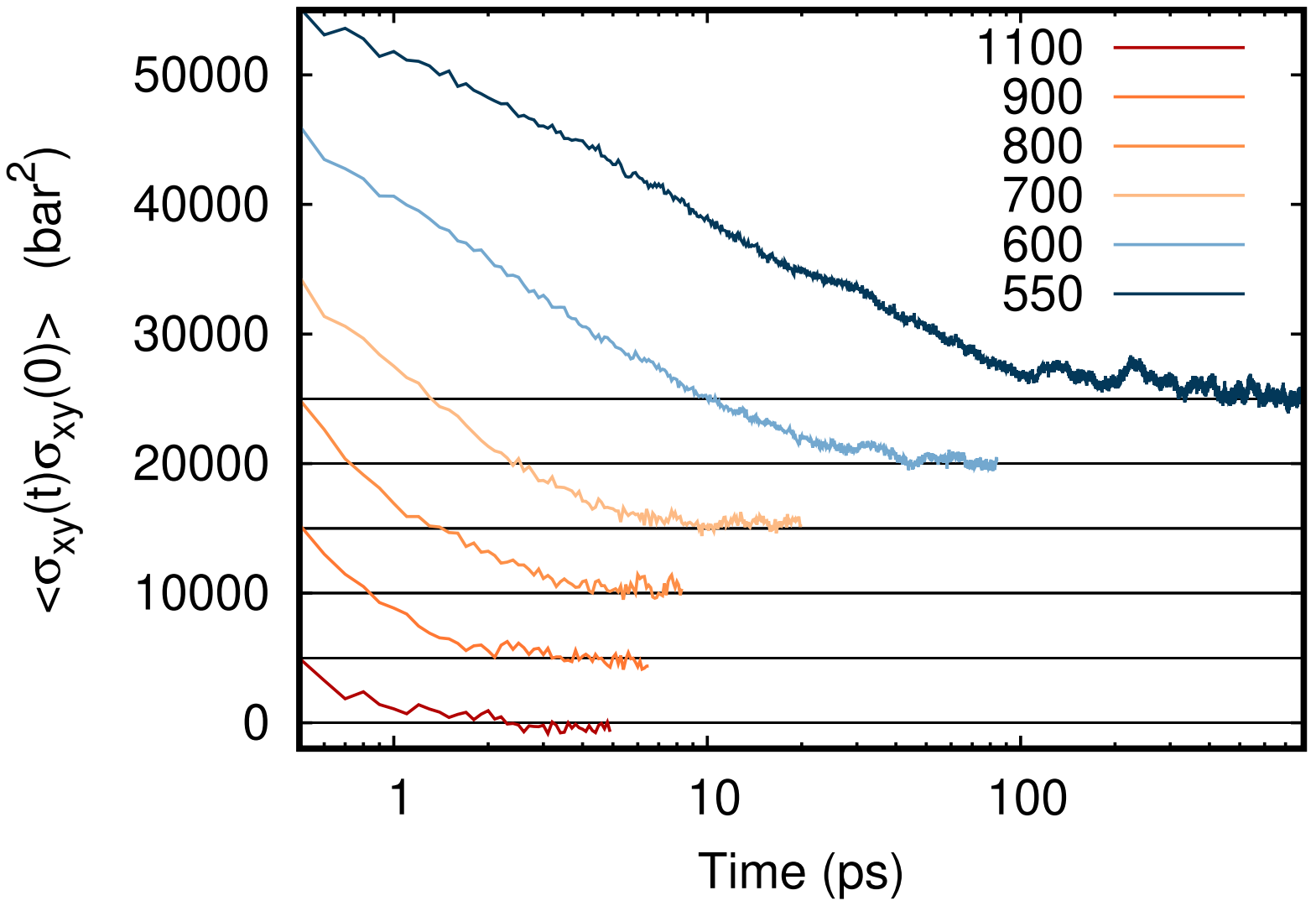

Figure S9. The self-correlation function $\langle\sigma_{xy}(t)\sigma_{xy}(0)\rangle$ entering in the Green-Kubo integral for the evaluation of the viscosity (see article) for a few representative temperatures. The curves are displaced vertically by 5000 $bar^2$ each, to improve the readability of the figure.

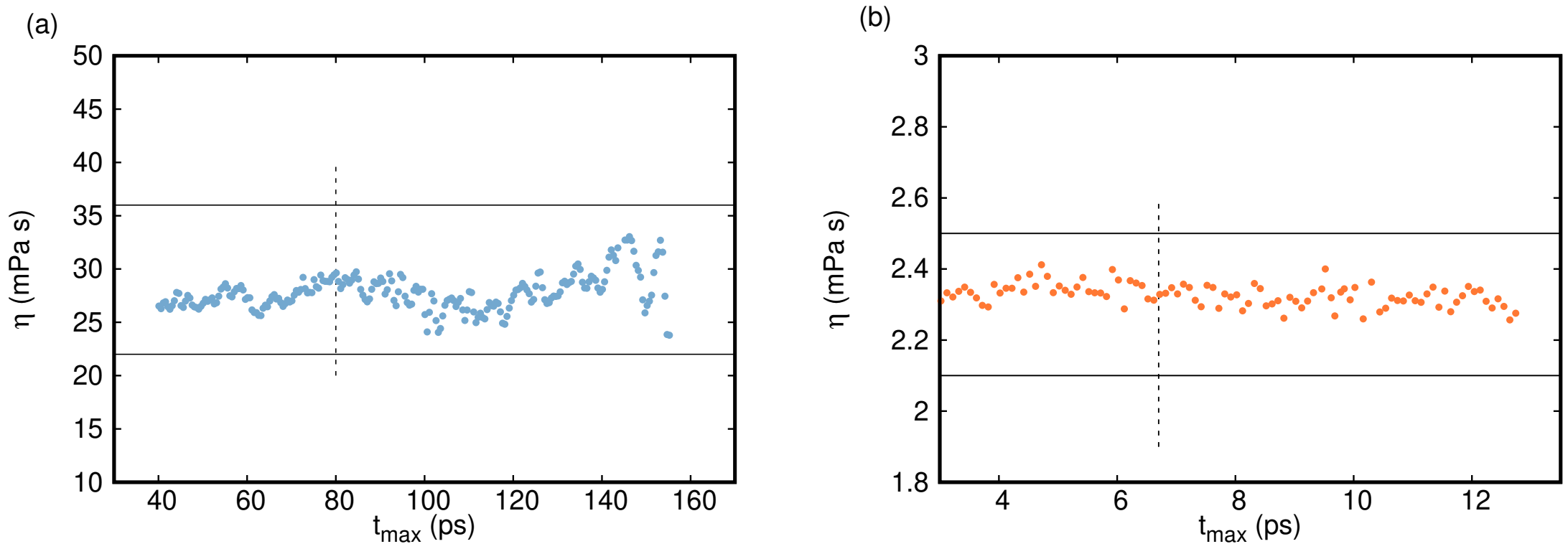

Figure S10. Dependence of our evaluation of $\eta$ on the choice of the maximum integration time $t_{max}$ in the GK formula (Eq. (1) in the article) at low (600 K, left panel) and high (900 K, right panel) temperatures. The horizontal lines refer to the highest and lowest estimation of $\eta$ given our error-bars, while the vertical dashed line corresponds to our choice of $t_{max}$.

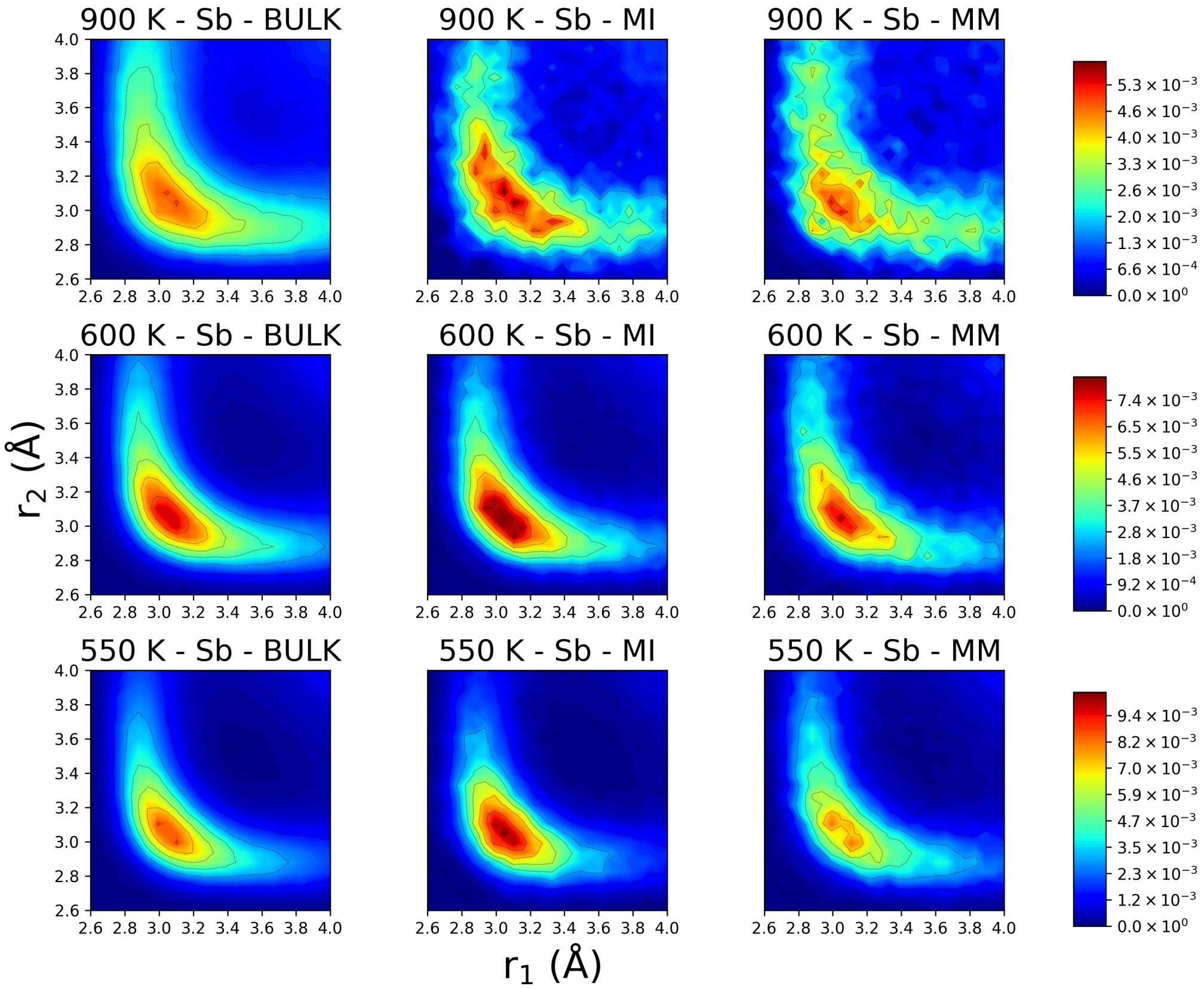

Figure S11. ALTBC function for Sb atoms at (top panels) 900 K (central panels) 600 K and (lower panels) 550 K for the bulk (left panels), the two largest MI clusters (central panels) and the two largest MM clusters (right panels).

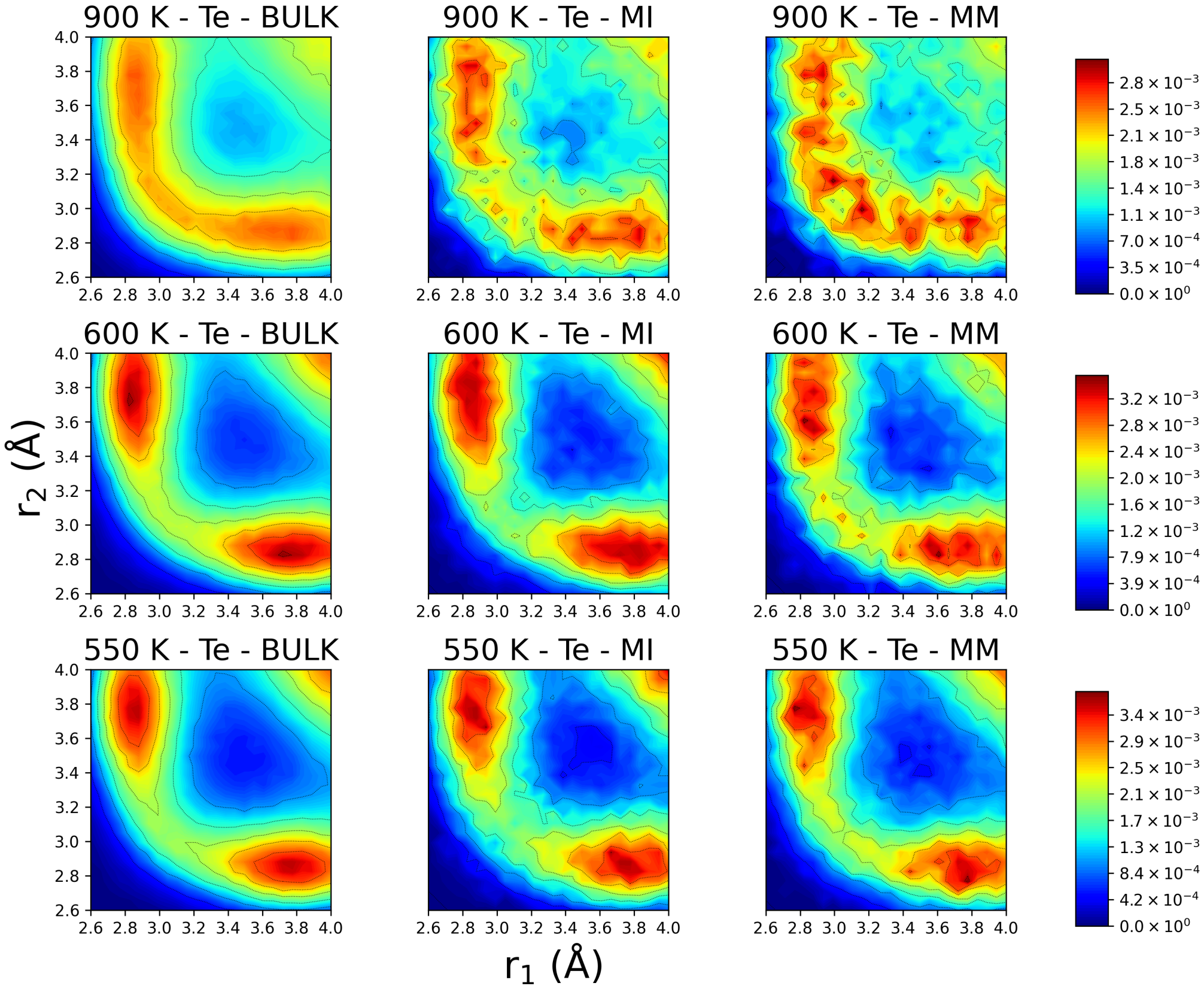

Figure S12. ALTBC function for Te atoms at (top panels) 900 K (central panels) 600 K and (lowe panels) 550 K for the bulk (left panels), the two largest MI clusters (central panels) and the two largest MM clusters (right panels).

| Temperature (K) | simulation ensemble | density (atom/Å$^3$) | simulation time (ns) | $t_{max}$ (ps) | $\eta$ (mPa s) | $D_{Ge}$ ($10^{-5}$ cm$^2$/s) | $D_{Sb}$ ($10^{-5}$ cm$^2$/s) | $D_{Te}$ ($10^{-5}$ cm$^2$/s) |
|---|---|---|---|---|---|---|---|---|
| 1200 | NVT | 2.993 | 6.6 | 3 | $\eta$=1.15 $\pm$ 0.07 * | | | |
| 1203 | NVE | 2.993 | 0.12 | | | 7.84 | 7.36 | 6.62 |
| 1200 | NVT | 3.075 | 6.6 | 3.2 | $\eta$=1.20 $\pm$ 0.08 * | | | |
| 1208 | NVE | 3.075 | 0.12 | | | 7.29 | 6.95 | 6.19 |
| 1150 | NVT | 3.009 | 8.8 | 4 | $\eta$=1.20 $\pm$ 0.06 * | | | |
| 1150 | NVE | 3.009 | 0.12 | | | 6.64 | 6.45 | 5.83 |
| 1150 | NVT | 3.075 | 8.8 | 4.2 | $\eta$=1.31 $\pm$ 0.08 * | | | |
| 1149 | NVE | 3.075 | 0.12 | | | 6.44 | 6.11 | 5.42 |
| 1100 | NVT | 3.023 | 8.8 | 5 | $\eta$=1.34 $\pm$ 0.08 * | | | |
| 1107 | NVE | 3.023 | 0.2 | | | 6.12 | 5.76 | 5.12 |
| 1100 | NVT | 3.075 | 8.8 | 5 | $\eta$=1.39 $\pm$ 0.07 * | | | |
| 1104 | NVE | 3.075 | 0.2 | | | 5.72 | 5.38 | 4.81 |
| 1050 | NVT | 3.039 | 8.8 | 5 | $\eta$=1.52 $\pm$ 0.06 * | | | |
| 1043 | NVE | 3.039 | 0.2 | | | 5.32 | 4.93 | 4.17 |
| 1050 | NVT | 3.075 | 8.8 | 5.6 | $\eta$=1.6 $\pm$ 0.1 * | | | |
| 1053 | NVE | 3.075 | 0.2 | | | 5.28 | 4.53 | 4.09 |
| 1000 | NVT | 3.052 | 8.8 | 6 | $\eta$=1.7 $\pm$ 0.1 * | | | |
| 1000 | NVE | 3.052 | 0.2 | | | 4.72 | 3.98 | 3.74 |
| 1000 | NVT | 3.075 | 8.8 | 6.2 | $\eta$=1.7 $\pm$ 0.08 * | | | |
| 1006 | NVE | 3.075 | 0.2 | | | 4.40 | 3.88 | 3.63 |
| 950 | NVT | 3.063 | 8.8 | 6 | $\eta$=1.8 $\pm$ 0.1 * | | | |
| 957 | NVE | 3.063 | 0.2 | | | 3.87 | 3.64 | 3.32 |
| 950 | NVT | 3.075 | 8.8 | 6.4 | $\eta$=1.9 $\pm$ 0.1 * | | | |
| 952 | NVE | 3.075 | 0.2 | | | 3.73 | 3.64 | 3.13 |
| 900 | NVT | 3.075 | 8.8 | 6.6 | $\eta$=2.3 $\pm$ 0.2 * | | | |
| 902 | NVE | 3.075 | 0.2 | | | 3.05 | 2.66 | 2.40 |
| 850 | NVT | 3.075 | 8.8 | 8 | $\eta$=2.7 $\pm$ 0.2 * | | | |
| 846 | NVE | 3.075 | 0.6 | | | 2.35 | 2.29 | 1.82 |

| Temperature (K) | simulation ensemble | density (atom/Å$^3$) | simulation time (ns) | $t_{max}$ (ps) | $\eta$ (mPa s) | $D_{Ge}$ ($10^{-5}$ cm$^2$/s) | $D_{Sb}$ ($10^{-5}$ cm$^2$/s) | $D_{Te}$ ($10^{-5}$ cm$^2$/s) |
|---|---|---|---|---|---|---|---|---|
| 800 | NVT | 3.075 | 8.8 | 8.4 | $\eta$=3.5 ± 0.5 * | | | |
| 794 | NVE | 3.075 | 0.4 | | | 1.81 | 1.62 | 1.43 |
| 750 | NVT | 3.075 | 11 | 17.2 | $\eta$=5.1 ± 0.7 * | | | |
| 751 | NVE | 3.075 | 0.4 | | | 1.29 | 1.13 | 1.0 |
| 700 | NVT | 3.075 | 11 | 19.2 | $\eta$=6.9 ± 0.9 * | 0.87 | 0.75 | 0.66 |
| 698 | NVE | 3.075 | 0.4 | | | 0.866 | 0.749 | 0.659 |
| 675 | NVT | 3.075 | 11 | 24 | $\eta$=9.5 ± 0.9 * | | | |
| 674 | NVE | 3.075 | 0.4 | | | 0.652 | 0.558 | 0.495 |
| 650 | NVT | 3.075 | 13.2 | 28 | $\eta$=13 ± 2 * | | | |
| 647 | NVE | 3.075 | 0.6 | | | 0.525 | 0.426 | 0.398 |
| 625 | NVT | 3.075 | 13.2 | 76 | $\eta$=13 ± 2 * | | | |
| 621 | NVE | 3.075 | 0.55 | | | 0.341 | 0.285 | 0.248 |
| 600 | NVT | 3.075 | 13.2 | 84 | $\eta$=29 ± 7 * | | | |
| 597 | NVE | 3.075 | 0.6 | | | 0.237 | 0.194 | 0.168 |
| 590 | NVT | 3.075 | 17.6 | 140 | $\eta$=39 ± 7 * | | | |
| 588 | NVE | 3.075 | 1.0 | | | 0.178 | 0.143 | 0.127 |
| 580 | NVT | 3.075 | 17.6 | 180 | $\eta$=57 ± 8 * | | | |
| 580 | NVE | 3.075 | 2.0 | | | 0.126 | 0.109 | 0.092 |
| 570 | NVT | 3.075 | 18.0 | 199 | $\eta$=90 ± 12 * | | | |
| 573 | NVE | 3.075 | 2.0 | | | 0.108 | 0.083 | 0.070 |
| 560 | NVT | 3.075 | 18.0 | 300 | $\eta$=153 ± 34 * | | | |
| 562 | NVE | 3.075 | 2.0 | | | 0.086 | 0.071 | 0.061 |
| 550 | NVT | 3.075 | 18.0 | 800 | $\eta$=218 ± 46 * | | | |
| 551 | NVE | 3.075 | 2.0 | | | 0.065 | 0.048 | 0.042 |

Table SI. Synoptic table of the different simulations (see article) with temperature, simulation ensemble, density, simulation time, maximum time $t_{max}$ used in the Green-Kubo integral for the viscosity, the resulting viscosity and specie resolved diffusion coefficients. The * indicates that $\eta$ is obtained by block averaging over several simulations under the same conditions (see article), the corresponding simulation time is the overall time summed over the independent simulations.